# THE [CII] 158 μm LINE DEFICIT IN ULTRALUMINOUS INFRARED GALAXIES REVISITED[1]

M. L. Luhman[2,3], S. Satyapal[4,5,6], J. Fischer[2], M. G. Wolfire[7], E. Sturm[8], C. C. Dudley[2,9], D. Lutz[8] & R. Genzel[8]

## ABSTRACT

We present a study of the [CII] 157.74 μm fine-structure line in a sample of 15 ultraluminous infrared (IR) galaxies (IR luminosity $L_{IR} \geq 10^{12}$ $L_\odot$; ULIRGs) using the Long Wavelength Spectrometer (LWS) on the *Infrared Space Observatory* (ISO). We confirm the observed order of magnitude deficit (compared to normal and starburst galaxies) in the strength of the [CII] line relative to the far-infrared (FIR) dust continuum emission found in our initial report (Luhman et al. 1998; Paper I), but here with a sample that is twice as large. This result suggests that the deficit is a general phenomenon affecting 4 of 5 ULIRGs. We present an analysis using observations of generally acknowledged photodissociation region (PDR) tracers ([CII], [OI] 63 and 145 μm, and FIR continuum emission), which suggests that a high ultraviolet flux $G_0$ incident on a moderate density n PDR could explain the deficit. However, comparisons with other ULIRG observations, including CO (1-0), [CI] (1-0), and 6.2 μm polycyclic aromatic hydrocarbon (PAH) emission, suggest that high $G_0$ / n PDRs *alone* cannot produce a self-consistent solution that is compatible with all of the observations. We propose that non-PDR contributions to the FIR continuum can explain the apparent [CII] deficiency. Here, unusually high $G_0$ and/or n physical conditions in ULIRGs as compared to those in normal and starburst galaxies are not required to explain the [CII] deficit. Dust-bounded photoionization regions, which generate much of the FIR emission but do not contribute significant [CII] emission, offer one possible physical origin for this additional non-PDR component. Such environments may also contribute to the observed suppression of FIR fine-structure emission from ionized gas and PAHs, as well as the warmer FIR colors found in ULIRGs. The implications for observations at higher redshifts are also revisited.

Subject headings: galaxies: high-redshift — galaxies: ISM — galaxies: starburst — infrared: galaxies — ISM: lines and bands

[1] Based on observations with ISO, an ESA project with instruments funded by ESA Member States (especially the PI countries: France, Germany, the Netherlands, and the United Kingdom) with the participation of ISAS and NASA
[2] Naval Research Laboratory, Remote Sensing Division, Code 7217, Washington, DC 20375
[3] Current address: Institute for Defense Analyses, 4850 Mark Center Drive, Alexandria, VA 22311-1882
[4] Goddard Space Flight Center, Code 685, Greenbelt, MD 20771
[5] Current address: George Mason University, Department of Physics and Astronomy, Fairfax, VA 22030-4444
[6] Presidential Early Career Award Scientist
[7] University of Maryland, Department of Astronomy, College Park, MD 20742
[8] Max-Planck-Institute für Extraterrestrische Physik, Postfach 1603, D-85740, Garching, Germany
[9] NRL/NRC Research Associate

## 1. INTRODUCTION

In a previous study (Luhman et al. 1998; hereafter Paper I), we reported measurements of the 157.74 μm $^2P_{3/2} - {^2P_{1/2}}$ fine structure line of $C^+$ in a sample of seven ultraluminous infrared (IR) galaxies (IR luminosity $L_{IR} > 10^{12}$ $L_\odot$; ULIRGs) using the Long Wavelength Spectrometer (LWS) on the *Infrared Space Observatory* (ISO). The observations presented in Paper I identified for the first time a [CII] deficit in ULIRGs relative to the far-infrared (FIR) dust continuum, whereby the [CII] flux is about an order of magnitude lower than that seen from nearby normal and starburst galaxies. As an important coolant of atomic and molecular interstellar gas exposed to ultraviolet (UV) photons, commonly referred to as photodissociation regions or PDRs (Tielens & Hollenbach 1985), the [CII] 158 μm line is often the single brightest line in the spectrum of most galaxies, providing 0.1%–1% of the total FIR luminosity (see, for example, Stacey et al. 1991 and references therein). However, based on the Paper I observations, the [CII] contribution in ULIRGs is more typically 0.01%–0.1% of the total FIR luminosity.

The unexpected low [CII] luminosity result could have important consequences for future spectroscopic surveys of protogalaxies from the ground or space (e.g., Stark 1997; Blain et al. 2000). The ULIRGs in this study are local examples of merging galaxies (see Sanders & Mirabel 1996 for a review). Since protogalaxies at high redshift (z = 1–10) are believed to form from the merging of smaller structures (e.g., Mobasher, Rowan-Robinson, & Georgakakis 1996; van den Bergh et al. 1996), they could resemble the nearby ULIRGs. As the brightest line in the spectrum of most galaxies and particularly bright in low metallicity galaxies (Madden 2000), the [CII] line is a potentially powerful spectroscopic tracer of protogalaxies, where for redshifts z = 2–6, the [CII] 158 μm line shifts to sub-millimeter wavelengths. Predictions of the detection rate of the [CII] line as a signpost of high-redshift galaxies must, however, take into account the [CII] deficit found in Paper I.

In Paper I, we did not have sufficient data in hand to explore in depth the source of the [CII] deficiency in ULIRGs. However, since the publication of Paper I, we have obtained additional FIR spectra – increasing our sample size to 15 ULIRGs – to confirm the [CII] deficit and readdress its origin. In some cases, we combined our added spectra with archival [CII] data from other ISO programs. In addition to the added FIR spectra, several new studies have been performed which provide useful clues regarding the nature of the [CII] emission in ULIRGs. Satyapal et al. (2003) and S. Satyapal et al. (in preparation) performed detailed analyses of the ISO FIR spectra of the luminous galaxy, Arp 299, and applied these analyses to a discussion of the FIR line emission in ULIRGs in general; Fischer et al. (1999, 2001) reported on the emission and absorption line characteristics of the FIR spectra of a small sample of IR-bright galaxies, including the ULIRG Arp 220, and modeled the photoionization of dust-bounded nebulae; and Genzel et al. (1998), Lutz et al. (1998), Rigopoulou (1999), Dudley (1999), Tran et al. (2001), and C. C. Dudley et al. (in preparation) studied the polycyclic aromatic hydrocarbon (PAH) emission in luminous IR galaxies. Several of these studies tend to suggest the probable importance of an additional component of dust emission not traced by PDR line diagnostics.

In related studies, Malhotra et al. (2001) analyzed the FIR spectra of a sample of 60 normal, star-forming galaxies. For the same sample of galaxies, Helou et al. (2000) compared the [CII] line emission to the 5–10 μm band aromatic feature emission and found evidence for a decrease in the importance of PAHs and small grains relative to large grains with increasing radiation field both in gas heating and in dust cooling. For those FIR-bright galaxies in the Malhotra et al. sample that display a [CII] deficit, Malhotra et al. attributed the decrease in the [CII] to FIR flux to PDRs with a high ratio of incident UV radiation $G_0$ to gas density n based on the values of $G_0$ and n derived from plane parallel PDR models. In a similar study, however, Negishi et al (2001) found that the ratio of $G_0$ to n remains constant in their sample of 34 galaxies and surmised instead that galaxies with low ratios of [CII] to FIR flux could be interpreted in terms of a decrease in the ionized component of the [CII] emission relative to the FIR intensity. Using the CO (1-0) and [CI] (1-0) line emission, several authors (Genzel & Cesarsky 2000; Gerin & Phillips 2000) have also recently derived PDR physical conditions for ULIRGs. The derived physical parameters for these various PDR studies vary significantly and show that, depending upon which lines are used, PDR analyses can give different results.

In this paper, we readdress the [CII] deficit in ULIRGs, combining the results of these recent ULIRG studies with our own ISO LWS observations of [CII] as well as archival data from other ISO programs for an enlarged sample of ULIRGs. Along with the [CII] data, we also report [OI] 63 µm measurements and [OI] 145 µm upper limits in a subset of the sample ULIRGs. Like the [CII] 158 µm line, the [OI] 63 µm line and, to a much lesser degree, the [OI] 145 µm line are strong cooling lines in Galactic and extragalactic PDRs (e.g., Tielens & Hollenbach 1985; Wolfire, Tielens, & Hollenbach 1990). The combined [CII] and [OI] data set, along with the FIR emission, allows us to derive estimates of $G_0$ and n using the predictions of PDR models and assuming that the [CII], [OI], and FIR emission each emanate from the same effective PDR surface area. In order to treat the PDR emission in a self-consistent way, we compare the [CII], [OI], and FIR emission to other PDR diagnostics, namely the CO (1-0), CI (1-0), and 6.2 µm PAH flux, and determine the proper application of one-dimensional, plane-parallel PDR models to ULIRG observations. Our analysis suggests the presence of a contributing component of non-PDR FIR emission to the total FIR luminosity in ULIRGs. We discuss the consequences of this result on the usefulness of [CII] as a tracer of high-redshift protogalaxies.

## 2. OBSERVATIONS AND RESULTS

In Table 1, we summarize the ULIRG observations. The ISO programs from which we extracted [CII] data include the ISO LWS consortium "Central Programme" on IR-bright galaxies (IRBGALS and IRBGALS2), which provided the data for Paper I; the ISO OH megamasers program (OHMEGA); our follow-up ISO discretionary-time program on the [CII] emission in ULIRGs (CII_ULIRG); and the Max-Planck-Institute für Extraterrestrische Physik extragalactic program (MPEXGAL2). We also extracted archival data from several other ISO programs (LWS_PV1, IRB_HD, SF_GLX_A), which were found to contain [CII] spectra. In general, the IRBGALS and CII_ULIRG programs contain observations of the [CII] 158 µm line only, whereas the MPEXGAL2 program contains both [CII] 158 μm and [OI] 63 μm observations. These line observations used the ISO LWS medium resolution (Δλ = 0.6 μm at 158 μm or $R \sim 200$) grating mode (Clegg et al. 1996). In the cases of Mkn 231, Mkn 273, Arp 220, and IRAS 17208-0014, we obtained full 43−197 μm spectra, from which we extracted flux measurements of the [CII] 158 μm and [OI] 63 μm and 145 μm lines.

Pipeline processing (Pipeline 10) of the raw ISO LWS spectra calibrated the wavelength and flux, removed cosmic ray glitches, subtracted dark current, and corrected for instrumental responsivity variations (Swinyard et al. 1996). We performed subsequent data processing, including co-addition, line profile extraction, and line flux and uncertainty measurement, using the ISO Spectral Analysis Package (ISAP[9]; Sturm et al. 1998). For sources where multiple observations were taken from different programs, the line fluxes are the weighted means with propagated errors. Table 2 contains the enlarged sample of ULIRGs and their corresponding [CII] line fluxes or upper limits. The [CII] fluxes here using Pipeline 10 processing compare well with the Pipeline 5 fluxes reported in Paper I, although because of the co-addition and use of Pipeline 10, the errors have decreased on average. The present study increases the number of ULIRGs in our sample from seven to 15. (We note, however, that $L_{IR}$ is slightly less than $10^{12}$ $L_\odot$ for two galaxies in our sample, IRAS 19254-7245 and IRAS 23128-5919; these galaxies are not strictly defined as ULIRGs. To within the uncertainties of the $L_{IR}$ estimates, however, these two galaxies are indistinguishable from $10^{12}$ $L_\odot$ galaxies so we include them in our ULIRG sample.) Additionally, Table 2 lists the [OI] 63 μm and 145 μm line flux measurements and upper limits where available. The absolute flux calibration error is ±25% and was determined by comparing overlapping data in neighboring detectors from the full grating scans. All sources in this study were spatially unresolved, with the ISO beam (69" at 158 μm) encompassing all of the IR emission.

As shown in Table 2, the [CII] line has been detected in 12 of the 15 ULIRGs in our sample. Of the nine ULIRGs where we have 63 μm spectra, the [OI] 63 μm line is seen in absorption in Arp 220 (Fischer et al. 1999), and in emission in Mkn 231, Mkn 273, IRAS 17208-0014, IRAS 20551-4250, and IRAS 23365+3604. We measured [OI] 145 μm upper limits for Mkn 231, Mkn 273, Arp 220, and IRAS 17208-0014.

In Figure 1, we plot the [CII] line flux against the FIR flux for the ULIRGs in our sample and for an assortment of normal and starburst galaxies from previous studies (Paper I; Bradford et al. 1999; Colbert et al. 1999; Stacey et al. 1999; C. C. Dudley et al., in preparation; J. Fischer et al., in preparation). Galaxies are identified by name according to the key in the figure. The comparison sample of normal and starburst galaxies is limited to sources with ISO 6.2 μm spectroscopy where the beam includes the whole galaxy or the same region as the LWS beam. Among the comparison sample of galaxies, four galaxies – NGC 6240, NGC 4945, NGC 1068, and Circinus – with optical, X-ray, or gamma ray signatures of active galactic nuclei (AGN) are included (e.g., Done, Madejski, & Smith 1996; Vignati et al. 1999).

We find that the enlarged sample of ULIRGs presented here confirms and reinforces the result presented in Paper I; that is, relative to the FIR continuum, the [CII] flux from ULIRGs is approximately an order of magnitude lower than that seen from nearby normal and starburst galaxies (as well as the AGN). Of the 15 ULIRGs, only three objects − IRAS 15206+3342, IRAS 19254-7245, and IRAS 23128-5919 − display values of $F_{[CII]} / F_{FIR}$ within the range 0.1%−1% indicative of virtually all normal and starburst galaxies (Crawford et al. 1985; Stacey et al. 1985, 1991; Wright et al. 1991). The $F_{[CII]}$ to $F_{FIR}$ ratio for the remaining 80% of the sample ULIRGs is less than 0.1%, with a mean value of 0.041% excluding sources with upper limits. The median of all ULIRGs in our sample is 0.039%, where for those ULIRGs with [CII] upper limits we use the 2σ value. This median value is 13% of the median $F_{[CII]}$ to $F_{FIR}$ ratio (0.31%) for the 60 normal galaxies in the Malhotra et al. (2001) sample, again using the 2σ value for those normal galaxies with [CII] upper limits only. [The ratio of the average value of our sample (0.062%) and the Malhotra et al. sample (0.31%) is 20%.] For both the Malhotra et al. sample and our sample, Arp 220 shows the lowest detected $F_{[CII]}$ to $F_{FIR}$ ratio of 0.021%. NGC 4418 has the lowest upper limit of 0.0051% (Malhotra et al. 2001) of any galaxy in either sample. It is of interest to note that NGC 4418 is believed to harbor a buried AGN based on its mid-IR spectral properties (Roche et al. 1986; Dudley & Wynn-Williams 1997; Spoon et al. 2001).

We note that our convention as given in Table 2 for calculating the FIR flux, chosen for consistency of comparison with previous extragalactic studies of the [CII] to FIR emission (e.g., Stacey et al. 1991; Lord et al. 1996, Malhotra et al. 1997), overestimates the actual value of $F_{[CII]} / F_{FIR}$, since the FIR flux estimate accounts for the FIR emission only between 42.5 µm and 122.5 µm (Helou et al. 1988). The correction factor required to account for the FIR flux longward of 122.5 µm is typically in the range of 1.4–1.8 (Sanders & Mirabel 1996). Thus, the actual [CII] to FIR ratio is slightly less than presented in Figure 1. For some of the extended normal and starburst galaxies for which the IRAS Faint Source Catalog (FSC) may have missed some of the FIR emission (Rice et al. 1988), we measured using ISAP the ISO LWS fluxes, again from 42.5 to 122.5 µm to match the IRAS waveband, and plotted these values in Figure 1. For some of the more extended normal and starburst galaxies, we find the ISO fluxes to be as much as 40% higher than the IRAS FSC fluxes. For galaxies small with respect to the LWS beam, namely all ULIRGs and also Arp 299, NGC 6240, NGC 7552, NGC 5713, and NGC 3256, we used the IRAS FSC for the FIR flux. For these ULIRGs and compact non-ULIRGs, the ISO LWS and IRAS FSC flux measurements show good agreement with an average ISO to IRAS FIR (42.5 to 122.5 µm) flux ratio of $1.10 \pm 0.09$.

For comparison, we plot in Figure 2 the 6.2 µm PAH feature flux (C. C. Dudley et al., in preparation) versus the [CII] emission. The galaxies plotted are the same as in Figure 1, with the exception of IRAS 22491-1808, which is not plotted. Neither [CII] nor the 6.2 µm PAH feature were detected for this ULIRG and therefore its upper limits do not constrain its position relative to the band shown (dashed lines). The 6.2 µm emission feature arises from ionized PAHs (Allamandola, Hudgins, & Sandford 1999). We take the 6.2 µm PAH data from C. C. Dudley et al. (in preparation), who measure the PAH emission from ISO PHOT-S spectra and ISO CAM CVF spectra integrated over the LWS profile-weighted beam for a flux limited sample of luminous IR galaxies. For three of the extended galaxies (Arp 299, NGC 253, and Circinus), only ISO CAM data over a 48" × 48" image are available (smaller than the 69" LWS aperture), but we infer from larger broadband ISO CAM images that these images include most of the flux. Dudley et al. chose the 6.2 µm feature because it is relatively isolated from the nearby 7.7/8.7 µm PAH feature blend whose measured strength can be affected by the 9.7 µm silicate absorption feature. Estimates of the 7.7 µm feature flux must therefore account for the effects of silicate absorption, which is especially difficult for redshifted sources with PHOT-S, which only extends to 11.5 μm in the observed frame.

In Figure 3, we display the distribution of the 6.2 μm PAH to [CII] 158 μm line ratio in histogram format for the ULIRGs and for the normal and starburst galaxies (no galaxies with AGN signatures are included in the comparison sample). Both Figures 2 and 3 show that, in contrast to the [CII] to FIR deficit found in ULIRGs, the PAH to [CII] ratio is similar in ULIRGs and the comparison sample, i.e., both [CII] and the 6.2 μm PAH feature display a proportionally similar deficit with respect to FIR in ULIRGs and normal/starburst galaxies. Using PHOT-S, Lutz et al. (1998) and Rigopoulou et al. (1999) find the 7.7 μm PAH to IRAS 60 μm ratio emission to be approximately three times weaker in ULIRGs than in starbursts. For these studies, the starburst galaxies are mostly relatively nearby sources, where in Rigopoulou et al., 11 out of 14 starburst galaxies are within cz $\leq$ 560 km s$^{-1}$. Since the aperture of the PHOT-S observations was 24" $\times$ 24", while the IRAS 60 μm fluxes likely arise in a significantly larger area in the starburst galaxies, the factor of three difference in the PAH to IRAS 60 μm ratio is likely an underestimate. Here, we attempted to avoid this problem by using ISO CAM CVF measurements for the nearby starburst galaxies, from which it is possible to measure only the PAH emission corresponding to the LWS measurement. Furthermore, for the reasons stated above, we used the 6.2 μm feature rather the 7.7 μm feature used by Rigopoulou et al. Thus, we believe that the similarity between ULIRGs and starbursts in the PAH to [CII] ratios shown here is unbiased in these important respects.

## 3. DISCUSSION

In Paper I, we discussed three possible explanations for the [CII] deficit: (1) self-absorbed or optically thick [CII] emission; (2) saturation of the [CII] emission in a photodissociated gas with a high ratio of incident UV flux $G_0$ to n; or (3) the presence of a soft UV radiation field suggesting, for example, an older stellar population or absorption of UV in dusty photoionization regions. Based on the [CII] observations alone, we could not make a definitive case in favor of only one of the three proposed explanations, although self-absorbed or optically thick [CII] emission seemed the least likely explanation for the [CII] deficit. Using the added ISO [CII] and [OI] observations, we explore these and other possible sources for the [CII] deficit not posed in Paper I.

### 3.1. Saturation of [CII]: High $G_0$ / n PDRs

For PDRs with a high ratio of $G_0$ to n ($G_0$ / n $\geq$ 1–10 cm$^3$), theoretical PDR models show that [CII] saturates relative to FIR such that the ratio of $F_{[CII]}$ to $F_{FIR}$ drops below values of $10^{-4}$ (e.g., Wolfire, Tielens, & Hollenbach 1990; Kaufman et al. 1999). Here, the high $G_0$ increases the positive charge of dust grains, thereby reducing the energy of ejected photoelectrons and the fraction of UV photon energy that is converted to gas heating. As a result, gas cooling via the [CII] line falls. (We note that another type of [CII] saturation occurs when the gas density n lies well above the [CII] critical density. We discuss this case further in section 3.2.1.)

Our observed [CII] line fluxes combined with the available [OI] observations allow us to measure n and $G_0$ for several sample ULIRGs and thus directly address the possibility of saturation of the [CII] emission in PDR gas. The contour lines shown in Figure 4 depict the expected behavior of [CII] 158 μm / [OI] 63 μm and ([CII] 158 μm + [OI] 63 μm) / $I_{FIR}$ as a function of n and $G_0$ for extragalactic PDRs. (The UV flux $G_0$ is normalized to the solar neighborhood radiation field, or $1.6 \times 10^{-3}$ erg cm$^{-2}$ s$^{-1}$; Habing 1968.) The contour plot is based on the predictions of the Kaufman et al. (1999) PDR code, which is an updated version of earlier works stemming from the study by Tielens & Hollenbach (1985). Among other updates to the input physics, the code includes the effects of gas heating associated with the photoelectric ejection of electrons from very small grains (<15 Å) and PAHs (Bakes & Tielens 1994). The standard PDR model of Kaufman et al. adopts the fractional gas phase abundances as measured by Sembach & Savage (1996; i.e., C/H = $1.4 \times 10^{-4}$, O/H = $3 \times 10^{-4}$). For an extragalactic PDR, approximately half of the observed IR dust continuum emission comes from clouds facing away from the observer that do not contribute to the [CII] 158 μm and [O I] 63 μm emission directed along the line of sight owing to optical depth effects (see Wolfire et al. 1990 and the appendix of this paper). This "excess" IR continuum emission is accounted for in Figure 4 (and 5) by setting $I_{FIR}$ equal to twice the theoretical continuum intensity emitted by a face-on PDR (or $I_{FIR} = 2 \times [2 \times 1.6 \times 10^{-3}\, G_0 / 4\pi] = 5.1 \times 10^{-4}\, G_0$ erg cm$^{-2}$ s$^{-1}$ sr$^{-1}$).

In order to use Figure 4 to measure n and $G_0$ for our sample of ULIRGs, the observed [CII], [OI], and FIR emission must arise predominantly from the observed ensemble of PDR surfaces. (Since the physical size of the beam at the distance to the source is large compared with a single PDR, the observed [CII], [OI], and FIR emission arises from many PDR surfaces separated in both physical and velocity space.) Some portion of the diagnostic line emission may arise from non-PDR gas. (For now, we neglect non-PDR sources for the FIR emission; we treat this case further in section 3.2.) For example, ionized gas in the form of classical HII regions or warm (T ~ 8 × 10$^3$ K) diffuse gas can emit [CII] radiation; likewise, shock-excited gas can generate significant [OI] emission. The fraction of non-PDR line emission should be subtracted before using Figure 4.

Although we are unable to estimate the fractional contribution from non-PDR gas for the sources studied, we argue that the contribution is less than that of the PDR component. One method to estimate the [CII] contribution from ionized gas is to rely on HII region modeling, as presented by Carral et al. (1994) and Satyapal et al. (2003). Carral et al. point out that extragalactic HII regions can contribute as much as 30%−50% of the total [CII] luminosity. Using the [NII] 122 μm line, which traces ionized gas in both HII regions and diffuse media, Malhotra et al. (2000) and Negishi et al. (2001) also attribute as much as 30%−50% of the [CII] emission to ionized gas. For the ULIRGs studied here, [NII] 122 μm measurements are not available. In the case of the prototypical luminous starburst galaxy Arp 299 ($L_{IR}$ ~ 4.8 × 10$^{11}$ $L_\odot$), with luminosity approaching that of the ULIRGs, Satyapal et al. find that the contribution from ionized gas to the [CII] luminosity is closer to 15% based on the relative [CII] and [NII] 122 μm line strengths in Arp 299. Satyapal et al. conclude that, compared to the warm diffuse gas, classical HII regions provide the bulk of the ionized gas contribution to the [CII] emission.

For the shock contribution to the [OI] emission, shock models (e.g., Hollenbach & McKee 1989) predict [OI] 63 μm to [CII] 158 μm ratios greater than three. Such ratios are roughly consistent with those seen in AGN and starbursts with a strong shock-excited component (e.g., Spinoglio & Malkan 1992; van der Werf et al. 1993). Based on the generally low ratio (~1) of [OI] to [CII] emission in the ULIRGs, however, we believe that the contribution to the [OI] emission from shock-excited gas is small.

In Figure 4, we plot five ULIRGs − Mkn 231, Mkn 273, IRAS 17208-0014, IRAS 20551-4250, and IRAS 23356+3604 − for which we have both [CII] 158 μm and [OI] 63 μm emission line measurements. For comparison, we also plot the objects in the Malhotra et al. (2001) sample of 60 normal and starburst galaxies. With the possible exception of Mkn 231, we see that the implied PDR gas density n (~100–$10^3$ cm$^{-3}$) is similar to that of the normal and starburst galaxies, i.e., the [CII] to [OI] 63 μm ratios for the ULIRGs are comparable to those of the normal and starburst galaxies shown. Although the derived PDR densities are comparable to the normal and starburst galaxies, they fall well below the interstellar gas densities implied by CO observations of the ULIRGs (Solomon et al. 1997). Whereas the typical PDR density of the ULIRGs plotted in Figure 4 is ~200 cm$^{-3}$, the interstellar gas densities measured by Solomon et al. range from ~700 cm$^{-3}$ (Mkn 231 and IRAS 23365+3604) to $1.3–1.6 \times 10^3$ (Mkn 273 and IRAS 17208-0014).

In Figure 4, the implied UV flux $G_0$ (~500–$10^3$) incident on the PDR gas is significantly higher than that of the normal and starburst galaxies by a factor of two to five. As mentioned, for PDRs with high $G_0$ to n ratios, the high UV flux increases grain charging and reduces the efficiency of photoelectric heating and thus [CII] / FIR falls (Wolfire, Tielens, & Hollenbach 1990; Kaufman et al. 1999). Correcting the [CII] and [OI] line fluxes for contributions from non-PDR sources to the [CII] and [OI] emission could alter the derived n and $G_0$ considerably, but we would not expect the $G_0$ to n ratio to decrease. For example, applying a reasonable correction to the [CII] flux to account for emission from ionized gas would move the ULIRG data points in Figure 4 down and to the left in such a way that both n and $G_0$ would increase by a similar amount. Likewise, lowering the [OI] flux by some correction factor to account for the shocked gas would move the ULIRG data points toward higher $G_0$ and lower n, thus potentially increasing $G_0$ / n. For large corrections in the [OI] emission of a factor of two or more, the derived $G_0 / n$, which ranges from one to five in Figure 4, would increase by a factor of a few.

A high $G_0 / n$ cause for the ULIRG [CII] deficit would agree with the conclusion of Malhotra et al. (1997, 2001), who suggest that high $G_0 / n$ PDRs (specifically high n and higher $G_0$) account for the [CII] flux deficiencies seen in the FIR-bright objects among their sample of normal galaxies. Malhotra et al. base their argument on the observed decrease in the [CII] to FIR ratio as a function of the 60 to 100 μm flux. The 60 to 100 μm emission from PDRs measures average temperatures of the large grain population and should therefore increase with increasing UV flux $G_0$. In Figure 5, we plot the Figure 1 data as a function of the 60 to 100 μm flux ratio. Here, the drop in [CII] to FIR flux with the 60 to 100 μm flux ratio shows that the full ULIRG sample data is consistent with the result found by Malhotra et al. We emphasize, however, that it is a decrease in [CII] / FIR with increasing $G_0 / n$, not with increasing $G_0$ alone, that points to a decrease in the photoelectric heating efficiency as the cause for the [CII] deficit. As pointed out by Negishi et al. (2001), we would not expect [CII] / FIR to decrease if both $G_0$ and n increase together. If we take Figure 4 at face value, then $G_0$ would seem to increase with respect to n in ULIRGs as compared to normal and starburst galaxies; thus, Figure 5 appears to be consistent with a high $G_0 / n$ origin for the [CII] deficiency.

Assuming that we are correct in applying the PDR models to the ULIRGs, we explore the degree to which uncorrected self-absorption in the [OI] 63 μm line could influence the location of the ULIRGs in Figure 4. Fischer et al. (1999) have reported such self-absorption in the [OI] 63 μm line in the prototypical ULIRG Arp 220. The upper limit on the [OI] 145 μm line places a strong constraint on the amount by which we can feasibly correct the [OI] 63 μm line for self-absorption. Unlike the ground-state [OI] 63 μm line which is seen in absorption in Arp 220, the [OI] 145 μm line should not be self-absorbed since it originates from an energy level that is 228 K above the ground state and should not be populated in cold or diffuse gas. If, for example, we increase the [OI] 63 μm line emission by a factor of two, Mkn 231 shifts to a location in Figure 4 as shown by the absorption vector corresponding to n ~ 700 $cm^{-3}$ and $G_0$ ~ $3 \times 10^3$. According to Figure 6, for this value of n and $G_0$, the [CII] 158 μm to [OI] 145 μm ratio is ~7, which corresponds to the measured lower limit on the [CII] to [OI] 145 μm ratio for Mkn 231. Any correction for [OI] 63 μm self-absorption greater than a factor of two would therefore yield a predicted [CII] to [OI] 145 μm smaller than ~7, in disagreement with the observed lower limit for this ratio.

When we apply this same constraint to the other ULIRGs in Figure 4 with [OI] 145 μm upper limits, we find that even small corrections for self-absorption lead to inconsistencies with the observations, i.e., the [OI] 63 μm line is likely not self-absorbed in these ULIRGs. For Mkn 273, the observed lower limit on the [CII] to [OI] 145 μm ratio is 18. According to Figure 6, a [CII] to [OI] 145 μm ratio of 18 matches the derived n (~200 $cm^{-3}$) and $G_0$ (~700) from Figure 4. Any upward correction to the [OI] 63 μm line flux would drive up n and $G_0$, thus dropping the predicted [CII] to [OI] 145 μm line ratio below the measured lower limit, again in disagreement with the observations. The same holds true for IRAS 17208-0014, the remaining ULIRG in Figure 4 with an observed [OI] 145 μm upper limit. Thus, with the possible exception of Mkn 231, [OI] self-absorption does not appear to affect the derived PDR gas densities and UV fluxes for the ULIRGs in Figure 4.

Irrespective of the potential effects of [OI] 63 μm self-absorption on the derived PDR parameters, the [OI] 145 μm line, together with the [CII] and FIR fluxes, constrains the derived n and $G_0$. For example, for Arp 220: $F_{[CII]} / F_{[OI]\ 145\ \mu m} \geq 11$, $F_{[CII]} / F_{FIR} = 2.09 \times 10^{-4}$; for Mkn 273: $F_{[CII]} / F_{[OI]\ 145\ \mu m} \geq 18$, $F_{[CII]} / F_{FIR} = 5.37 \times 10^{-4}$; and for IRAS 17208-0014: $F_{[CII]} / F_{[OI]\ 145\ \mu m} \geq 11$, $F_{[CII]} / F_{FIR} = 4.73 \times 10^{-4}$. The only PDR regimes with these line ratios that are in keeping with both Figures 6 and 7 are either (1) $G_0$ ~ (0.7–2) × $10^3$ and n ~ 100 $cm^{-3}$ or (2) $G_0$ ~ 1–10 and n ~ $10^5$ $cm^{-3}$. The former case would characterize a high $G_0$ / n PDR, but n ~ 100 $cm^{-3}$ is much lower than the derived average interstellar densities of ~$10^4$ $cm^{-3}$ for Arp 220, ~1300 $cm^{-3}$ for Mkn 273, and ~1600 $cm^{-3}$ for IRAS 17208-0014 from CO observations (Solomon et al. 1997; Scoville, Yun, & Bryant 1998). (As we have already noted, the observations and PDR models together significantly underestimate the PDR gas density for the ULIRGs in Figure 4 as compared to previous CO estimates of the interstellar gas density.) The density for the latter case is considerably higher, on the other hand, but values of $G_0$ ~ 1−10 and $G_0$ / n ~

$10^{-4}$ in Arp 220, Mkn 273, and IRAS 17208-0014 are highly unlikely given the high luminosity of these galaxies. Thus, we find that any reasonable PDR model that can account for the observed [CII], [OI] 63 and 145 µm, and FIR emission does not compare well with previous observations of the neutral/molecular ISM in ULIRGs.

Putting aside for the moment the disparity between the n and $G_0$ for ULIRGs derived from the PDR analysis above and from previous studies, we examine the issue of high $G_0$ / n PDRs and their effect, if any, on the observed similarity of the 6.2 µm PAH to [CII] ratio in ULIRGs and in normal and starburst galaxies seen in Figures 2 and 3. In order for high $G_0$ / n PDRs to characterize ULIRGs but not most normal/starburst galaxies, high $G_0$ / n PDRs must decrease together both the [CII] *and* PAH flux relative to FIR in ULIRGs, excluding for now other effects such as mid-IR extinction (discussed in section 3.4), which could act to reduce the PAH flux relative to [CII]. In Figure 8, we plot the ratio of the 6.2 µm PAH to [CII] flux versus the 60 to 100 µm ratio, thought to scale with $G_0$ in the PDR. (As with Figure 2, IRAS 22491-1808 is not included in this plot since neither its 6.2 µm PAH nor [CII] line emission have been detected.) We see that the PAH to [CII] ratio stays fairly constant over the entire range of 60 to 100 µm ratios characteristic of normal/starburst galaxies and ULIRGs. We note that two of the four AGN-signature galaxies are relatively low in 6.2 µm PAH to [CII] flux; this tendency in AGN, if real, may be due to an enhanced contribution from a non-PAH generating component (such as the narrow-line region), AGN processing of PAH molecules, or other AGN effects (Imanishi & Dudley 2000). Helou et al. (2000) also observed an implied constancy in the PAH to [CII] ratio for their ISO normal galaxies key project using broadband mid-IR photometry to measure the PAH emission. [We note that low metallicity dwarfs can display systematically weaker PAH to [CII] ratios (Madden 2000; Hunter et al. 2001).] Helou et al. (2000) explain the close association of PAH / [CII] with 60 to 100 µm color by assuming that the PAH emission measures the PDR photoelectric heating, and [CII] measures the gas cooling. In this scenario, the decrease in [CII] / FIR (Figure 5) and PAH / FIR (inferred from Figures 5 and 8) with increasing radiation field would suggest a decrease in the importance of PAHs and small grains relative to large grains in both gas heating and dust cooling.

Although [CII] as a measure of PDR gas cooling is well established via both theory and observations, the association between the 6.2 μm feature emission from ionized PAHs and the photoelectric process as a function of $G_0$ is not so definitive. High $G_0$ / n conditions causing low values of the [CII] to FIR ratio would not produce a correspondingly low 6.2 μm feature flux, unless a significant fraction of the grains responsible for the 6.2 μm feature are destroyed by these conditions. However, laboratory studies indicate that PAH sizes greater than 50 carbon atoms probably dominate the emission in this band (Hudgins & Allamandola 1999), while theoretical studies indicate that it is difficult to destroy PAHs with greater than 50 carbon atoms in the regimes of $G_0$ / n derived here (Allain, Leach, & Sedlmayr 1996a,b). According to Bakes & Tielens (1994), it would appear that grains up to at least 100 Å in size would have to be destroyed in order to lower the heating by a factor of ten. Recent ISO observations also show little change in the mid-IR PAH spectral shapes over a wide range of $G_0$ ($1-10^5$; Uchida et al. 2000), even though laboratory studies show significant changes in relative strengths of different features from ionized to neutral PAHs. Because of such results, it is difficult to unambiguously associate the 6.2 μm feature emission from ionized PAHs with the photoelectric process as a function of $G_0$, even though Figure 8 and the similar finding of Helou et al. suggest that PAHs closely track the photoelectric heating.

Before we conclude our discussion of high $G_0$ / n PDRs and the relative PAH and [CII] emission, we note another interesting result of the Helou et al. study. Helou et al. find that at larger values of the 60 to 100 μm flux ratio, the [OI] 63 μm emission overtakes [CII] in cooling luminosity. Helou et al. point out that this observation conflicts with the argument that PAH / [CII] is constant because it is tied to the photoelectric effect, since as $G_0$ increases, the ratio of ([CII] + [OI]) / PAH rises, suggesting an increase in the photoelectric heating efficiency of the PAHs, which contradicts the physics. One scenario devised by Helou et al. to explain the observations is a two-component model whereby the FIR emission comes from both an "active" star formation PDR component and a "quiescent" PDR component. The "quiescent" PDR component would dominate at low values of 60 μm / 100 μm and display high values of [CII] / [OI], PAH / FIR, and [CII] / FIR; the "active" component would dominate at high 60 μm / 100 μm and show low values of [CII] / [OI], PAH / FIR, and [CII] / FIR. Although we do not observe especially low values of [CII] / [OI], a two-component model for the FIR emission raises the possibility that the ratio of 60 μm / 100 μm as plotted in Figures 5 and 8 may not necessarily track $G_0$ in the region that gives rise to the bulk of the PAH and [CII] emission. Instead, another FIR component could contribute to the integrated 60 and 100 μm fluxes, producing warmer 60 μm / 100 μm colors and increasing the FIR flux relative to [CII] and PAH as observed in Figures 5 and 8. The existence of multiple components to the FIR flux would bring into question our application of the one-dimensional, plane-parallel PDR models presented in this section. In the following section, we explore in more detail the possible contribution of non-PDR components to the FIR emission and its bearing on the [CII] deficit.

### 3.2. Applicability of PDR models: Non-PDR FIR Components?

A key assumption underlies the one-dimensional, plane-parallel PDR analysis presented in section 3.1: the beam area filling factors for the [CII], [OI], and FIR emission are taken to be equal or nearly so. Simply put, the PDR analysis assumes that the bulk of the [CII], [OI], and FIR emission comes from the same region of PDR gas with roughly similar average PDR parameters and that no other interstellar components contribute significantly to the observed emission. Here, the beam area filling factor Φ refers to the ratio of the effective PDR surface area intercepted by the beam to the beam projected area. Multiple PDRs along a line of sight can contribute to the filling factor as long as the background emission can escape to enter the beam. This can be due to velocity shifts between PDR layers so that line emission is not blocked by foreground gas, or from optically thin line emission at the same velocity. For distant extragalactic sources where the beam encompasses the entire galaxy, filling factor effects may be important (e.g., Wolfire, Hollenbach, & Tielens 1989). If non-PDR components of line or continuum emission are present in ULIRGs or if multiple or varying PDR components with vastly different conditions are present then the PDR models implying high $G_0$ / n PDRs do not apply. Such scenarios could account for the measured deficit in [CII] relative to FIR, since once corrected for filling factor effects, the [CII] to FIR ratios in the ULIRGs may resemble the ratios in normal and starburst galaxies.

We can estimate the [CII] beam area filling factor $\Phi_{[CII]}$ in the four ULIRGs for which we estimated n and $G_0$ above by comparing the observed [CII] intensity to the predicted [CII] intensity for the derived n and $G_0$. Table 3 shows the observed [CII] intensities for Mkn 273, IRAS 17208-0014, IRAS 20551-4250, and IRAS 23365+3604. We derive the [CII] intensities from the observed fluxes as measured in a 69" beam. Along with the observed intensities, Table 3 gives the predicted [CII] intensities (Kaufman et al. 1999) for a unity beam area filling factor given n and $G_0$ as derived in the previous section. The ratio of the measured [CII] intensity to the predicted [CII] intensity then provides an estimate of the beam area filling factor for the observed [CII] (Table 3, column 4). We note that, although the beam area filling factors derived in this manner are rough estimates, this approach can be used to make a relative comparison of beam area filling factors for various emission line or continuum components, as illustrated in the following sections.

### 3.2.1. CO (1-0) Emission Component

Having derived $\Phi_{[CII]}$ for four ULIRGs in our sample, we now examine whether the filling factors of other PDR emission lines, and ultimately the FIR continuum, match those of [CII]. This filling factor analysis is in effect a consistency check on our assumptions that the bulk of the PDR line diagnostics and FIR emission come from the same regions of PDR gas with roughly similar average PDR parameters. To our knowledge, such an analysis is done here in detail for the first time. We begin with the 2.6 mm CO (1-0) emission. Kaufman et al. (1999) model the behavior of the CO (1-0) intensity as a function of n and $G_0$, again assuming unity beam area filling factor. Using the Kaufman et al. predictions and previous observations of CO (1-0), we can estimate the CO (1-0) beam area filling factor $\Phi_{CO}$ as we did with [CII].

Table 4 gives the observed CO (1-0) intensities for Mkn 273, IRAS 17208-0014, IRAS 20551-2450, and IRAS 23365+3604 from previous studies scaled to the 69" beam of our ISO [CII] data. Again, taking the predicted n and $G_0$ for these galaxies from section 3.1, we list in Table 4 the predicted CO (1-0) intensities according to Kaufman et al. (1999). The ratio of the measured to predicted CO (1-0) intensities yields the CO (1-0) beam area filling factor $\Phi_{CO}$. Comparing the CO and [CII] beam area filling factors, we find that $\Phi_{CO}$ is four to 11 times larger than $\Phi_{[CII]}$ (Table 4, column 5). In other words, within the ISO 69" beam, the effective surface area of the environment that gives rise to CO (1-0) emission is four to 11 times that of the gas that emits [CII].

Having established that the [CII] and CO filling factors differ significantly for the four ULIRGs in Tables 3 and 4, and possibly, for ULIRGs in general, the question now becomes: How does $\Phi_{[CII]}$ compare to the FIR beam area filling factor $\Phi_{FIR}$? We cannot use the approach above to independently derive $\Phi_{FIR}$ for comparison with $\Phi_{[CII]}$, since both the observed [CII] and FIR flux were used to derive the PDR n and $G_0$. Instead, to address this question, we refer to a review paper by Genzel & Cesarsky (2000). In Figure 8a of their paper, Genzel & Cesarsky plot the intensity ratio $I_{[CII]} / I_{FIR}$ against $I_{CO (1-0)} / I_{FIR}$ for a variety of galaxy types including ULIRGs. Using PDR model overlays (Wolfire, Hollenbach, & Tielens 1989) to provide a measure of $G_0$ and n, Genzel & Cesarsky find that the location of the ULIRGs in their plot immediately suggests that the [CII] deficit can be quantitatively understood if the gas densities exceed $10^5$ cm$^{-3}$, with $G_0 \sim 10^3$–$10^4$, typical of Galactic PDR/HII star forming regions. This finding assumes, however, that the [CII], CO, and FIR beam area filling factors are identical. In this proposed high density case where n lies well above the [CII] critical density ($3 \times 10^3$ cm$^{-3}$ for collisions with hydrogen – Flower & Launay 1977; Hayes & Nussbaumer 1984), the abundance of C$^+$ in the upper level is fixed by the atomic statistical weights. Furthermore, the column density of C$^+$ becomes insensitive to $G_0$ (Wolfire, Hollenbach, & Tielens 1989). Since the abundance and column density cease to rise with n and $G_0$, the emergent [CII] line intensity saturates, giving rise to a [CII] deficit with respect to the FIR emission.

The interpretation that the [CII] line arises in high density PDR/HII regions cannot adequately explain the PDR line emission for ULIRGs, however. In particular, according to Figure 6, we see that for densities of $10^5$ cm$^{-3}$ and $G_0 \sim 10^3$–$10^4$, the strength of the [OI] 145 μm line should be comparable to that of the [CII] line. The observed [OI] 145 μm line strengths in the ULIRGs, on the other hand, fall well below that of the observed [CII] line. Additionally, assuming no [OI] 63 μm self-absorption, the Kaufman et al. PDR models imply a [OI] 63 μm to [CII] ratio of 10–30 for densities of $10^5$ cm$^{-3}$ and $G_0 \sim 10^3$–$10^4$, which is much higher than observed. This mismatch with the [OI] observations implies again that PDR models, together with the assumption that [CII], CO (1-0), and FIR beam filling factors are similar, are incompatible with the observations.

We have shown that for the four ULIRGs for which we have both [CII] and unabsorbed [OI] 63 μm emission line fluxes, $\Phi_{CO}$ is four to 11 times larger than $\Phi_{[CII]}$. If we assume that $\Phi_{[CII]} = \Phi_{FIR}$, i.e., that the observed [CII] to FIR ratio is the intrinsic ratio unaffected by beam area filling factor effects, then the observed ratio of $I_{CO\,(1-0)} / I_{FIR}$ must be divided by a factor of four to 11 ($= \Phi_{CO} / \Phi_{[CII]}$ or $\Phi_{CO} / \Phi_{FIR}$) before plotting the ULIRG data in Figure 8a ($I_{[CII]} / I_{FIR}$ versus $I_{CO\,(1-0)} / I_{FIR}$) of Genzel & Cesarsky (2000). We find that such a correction, however, would result in only a factor of a few change in the derived $G_0$ and n, and these values would still remain inconsistent with the observed [CII] and [OI] 63 μm and 145 μm line ratios.

On the other hand, if $\Phi_{CO} = \Phi_{FIR}$ for the ULIRGs, then $\Phi_{[CII]}$ and $\Phi_{FIR}$ must differ, and the observed $I_{[CII]} / I_{FIR}$ is not the intrinsic ratio. Specifically, for the four galaxies in Tables 3 and 4, we would correct $I_{[CII]} / I_{FIR}$ by multiplying by $\Phi_{CO} / \Phi_{[CII]}$ ($= \Phi_{FIR} / \Phi_{[CII]}$ for $\Phi_{CO} = \Phi_{FIR}$) in Table 3, column 5. In this instance, the location of the ULIRGs in Figure 8a of Genzel & Cesarsky implies a gas density n $\sim 10^3$–$10^4$ cm$^{-3}$ and a UV flux $G_0 \sim 10^3$–$10^4$. Such derived PDR properties are typical of normal and starburst galaxies and would provide agreement between the observed [OI] 63 μm and 145 μm and [CII] line ratios and the PDR models. Furthermore, such a beam area filling factor correction would remove the [CII] deficit yielding a [CII] to FIR ratio closer to 0.1%. *Thus, we could account for the [CII] deficit in ULIRGs with $G_0$ and n parameters typical for starburst galaxies if the beam area filling factor for FIR is significantly larger than that for [CII], as it is for CO. Here, the PDR model comparisons that support high $G_0 / n$ PDRs in ULIRGs would not apply because the beam area filling factors for [CII] (and presumably [OI]) and FIR are not the same.* The presence of additional non-PDR components of FIR and CO emission would provide one possible means of achieving FIR and CO luminosity without the accompanying [CII] emission.

If the above analysis is correct, we should see agreement in the beam area filling factors for [CII] and CO in normal and starburst galaxies, in contrast to the ULIRGs. To test this condition, we performed the above beam area filling factor analysis for an assortment of normal and starburst galaxies for which there exists [CII] and [OI] data (using ISO data where available) and CO (1-0) data with comparable spatial resolution as the [CII] and [OI] observations. The [CII] and CO beam area filling factors for these galaxies are shown in Tables 3 and 4. In all cases, $\Phi_{CO} / \Phi_{[CII]}$ is near unity for the normal and starburst galaxies, in contrast to our finding for the ULIRGs (Table 4, column 5). [We note that for some normal galaxies, particularly early type galaxies, characterized by low star formation rates, the [CII] emission may arise primarily from the diffuse ISM, in which case, we would not expect $\Phi_{[CII]}$ and $\Phi_{CO}$ to agree (Pierini et al. 2001).] Thus, we conclude that, for ULIRGs, the beam area filling factor differences between [CII] and CO are real.

The presence of CO emission arising largely from a non-PDR component where there is no [CII] is consistent with CO observations of Arp 220 and Mkn 231 (e.g., Downes & Solomon 1998; Scoville, Yun, & Bryant 1997). These observations show that much of the CO emission arises from molecular gas that is not cospatial with the nuclei (the presumed sites of the FIR emission) as seen in the millimeter continuum. The suggestion that, like CO (1-0), the FIR emission arises in large part from a non-PDR component provides a compelling explanation for the [CII] deficit, although without the benefit of an FIR interferometer, conclusive evidence that this is the case is difficult to produce. Although the location of the [CII] emission is uncertain, we note that it is not unreasonable to suggest that the [CII] emission may be coincident with the PAH emission (discussed further in section 3.2.3), which Soifer et al. (2002) have found is somewhat extended and not fully coincident with the mid-IR continuum in Arp 220 and Mkn 273. This would tend to support the idea that the FIR and [CII] emission may not be cospatial.

*3.2.2. CI (1-0) Emission Component*

The 609 μm [CI] (1-0) line is another fine-structure line arising in PDRs for which there exists ULIRG observations. Again, these observations and the PDR models allow us to derive a [CI] beam area filling factor $\Phi_{[CI]}$ for comparison with $\Phi_{[CII]}$. According to one-dimensional PDR models (e.g., Tielens & Hollenbach 1985), in clouds exposed to UV radiation, carbon is mostly in ionized form to a depth $A_V \sim 1$ mag. Atomic carbon appears at an intermediate ($A_V \sim 1$ to 5 mag) depth, where $C^+$ has mostly recombined with electrons to form C and not all of the gas phase carbon is converted into CO. The [CI] line thus traces regions exposed to nonionizing UV radiation. Since FIR continuum radiation comes from regions exposed to both ionizing and nonionizing UV radiation, estimates of $\Phi_{[CI]}$ can place a lower limit on $\Phi_{FIR}$.

Gerin & Phillips (2000) measure the [CI] (1-0) emission in a sample of 12 galaxies, in which two galaxies are ULIRGs – Arp 220 and Mkn 231. In the case of Arp 220, we have good estimates of n and $G_0$ from previous studies, which we can use to determine $\Phi_{[CI]}$. Gerin & Phillips measure a [CI] (1-0) intensity of $2.16 \times 10^{-7}$ erg cm$^{-2}$ s$^{-1}$ sr$^{-1}$ scaled to the ISO 69" beam. CO observations of Arp 220 suggest a moderately dense interstellar medium with $n \sim 10^4$ cm$^{-3}$ (Solomon et al. 1997; Scoville, Yun, & Bryant 1997; Gerin & Phillips 1998). Based on the FIR continuum and hydrogen recombination line emission in Arp 220, Gerin & Phillips (1998) estimate the UV radiation field $G_0$ to be $4 \times 10^4$. According to Kaufman et al. (1999), the predicted [CI] (1-0) intensity assuming unity beam area filling factor is $5.1 \times 10^{-6}$ erg cm$^{-2}$ s$^{-1}$ sr$^{-1}$. The ratio of the observed [CI] intensity to the predicted intensity yields a [CI] (1-0) beam area filling factor $\Phi_{[CI]}$ of 0.042 for a 69" beam. Similarly, taking the [CII] intensity in Arp 220 ($9.9 \times 10^{-6}$ erg cm$^{-2}$ s$^{-1}$ sr$^{-1}$) and comparing with the [CII] intensity predictions of Kaufman et al. (1999) for a unity beam area filling factor, we derive a [CII] beam area filling factor $\Phi_{[CII]}$ of 0.01 for Arp 220, or $\Phi_{[CI]} / \Phi_{[CII]} \sim 4$. For comparison, using the unity beam area filling factor CO intensity as predicted by Kaufman et al., we estimate that $\Phi_{CO}$ equals 0.17 ($I_{CO\,(1-0)} = 2.05 \times 10^{-8}$ erg cm$^{-2}$ s$^{-1}$ sr$^{-1}$ corrected to a 69" beam – Radford, Solomon, & Downes 1991; Gerin & Philips 1998), or $\Phi_{CO} / \Phi_{[CII]} \sim 17$.

Thus, as with the CO (1-0) emission in Mkn 273, IRAS 17208-0014, IRAS 20551-4250, and IRAS 23365+3604, we see that the implied projected area of the [CI]-emitting region (and CO-emitting region) is significantly larger than that of the [CII]-emitting region in Arp 220. We contend, as stated above, that the value of $\Phi_{[CI]}$ places a crude lower limit on $\Phi_{FIR}$. Multiplying the observed value of $F_{[CII]} / F_{FIR}$ in Arp 220 by $\Phi_{FIR} / \Phi_{[CII]}$ can therefore account for most, if not all, of the [CII] deficit.

For the sake of argument, let us assume that the [CII] emission from PDRs is coextensive with the [CI] and FIR emission, i.e., that the [CII] is intrinsically depressed with respect to [CI] and FIR. Discounting opacity and extinction effects (discussed in section 3.4), PDR models should therefore apply, and a unique $G_0$ and n regime that explains the [CII], [CI], and FIR emission should exist. For the 12 galaxies in their sample, Gerin & Phillips (2000) plot [CII] / FIR and [CII] / [CI] versus [CI] / FIR on contours of n and $G_0$ from PDR model predictions. Most of the galaxies studied lie in the parameter space, $G_0 = 10^2$–$10^3$, n = $10^2$–$10^5$, typical of molecular clouds in the nuclei of nearby galaxies. However, Arp 220 and Mkn 231 – the only ULIRGs in the sample – lie outside of the parameter space depicted by the modeled contours, but in the direction of very low $G_0$ / n ($\leq 0.01$). Such values of $G_0$ / n are at least two orders of magnitude lower than the estimates obtained with the FIR fine-structure lines, again suggestive of filling factor discrepancies attributable to a non-PDR component(s).

### 3.2.3. 6.2 μm PAH Emission Component

Referring again to Figure 2, the depicted [CII] to PAH emission behavior provides compelling support for the influence of filling factor effects on the observed emission. If the above filling factor analysis is correct, we should measure normal line ratios when comparing [CII] with another species that is known to come from the same gas. Like [CII], the 6.2 μm PAH emission traces gas exposed to far-UV radiation (Tielens & Hollenbach 1999; Vigroux et al. 1999). In Figure 2, we see that in fact the relative [CII] to PAH emission remains unchanged when comparing the ULIRGs to the normal and starburst galaxies. We would expect the observed difference in the [CII] to FIR and 6.2 μm PAH to FIR ratios between ULIRGs and normal/starburst galaxies to be a normal byproduct of an additional FIR component in ULIRGs that is not found in less luminous galaxies. Soifer et al. (2002) have confirmed that the mid-IR continuum in Arp 220 and Mkn 273 is strongly absorbed, while showing for the first time that the 11.3 μm PAH emission is extended. If the absorbed mid-IR radiation is powering the FIR emission, then even an apparent overlap between the PAH emission and the FIR emission would not mean that they share the same power source. *Thus, within the ISO LWS beam, as with CO and [CI], there could likely exist sources of FIR emission lacking associated PAH and [CII] emission that, once taken into account, could explain the [CII] deficit.*

### *3.2.4. Implications for Derived PDR Parameters*

Although we are unable to derive precise corrected PDR parameters resulting from the inferred presence of a non-PDR FIR component, we note that such a correction would move the ULIRGs to the right in Figure 4, resulting in higher densities and lower $G_0$ values compared with the uncorrected values. Depending on the amount of the correction, this is more consistent with the higher densities thought to be present in ULIRGs (Solomon et al. 1997). In this case, the warm 60 to 100 μm colors of ULIRGs would be generated in the non-PDR component of the FIR emission. We can see from Figure 6 that parameter changes in these directions (lower $G_0$ and higher n) would not have a strong effect on the [CII] to [OI] 145 μm ratio.

### 3.3. Soft UV Radiation Field: Aging Starbursts

In Paper I, a soft UV radiation field was presented as a possible explanation for the [CII] deficit in the observed ULIRGs. A stellar population deficient in massive main sequence stars, either as a result of an aging starburst or an initial mass function with a low upper mass cutoff, could give rise to a soft UV field and, consequently, a small ratio of $F_{[CII]} / F_{FIR}$ (Spaans et al. 1994; Nakagawa et al. 1995; Malhotra et al. 2001). S. Satyapal et al. (in preparation) model the decrease in $L_{[CII]} / L_{FIR}$ as a function of starburst age. In their model, an ionization-bounded HII region and PDR interface surround a central starburst cluster, with the location of the PDR gas lying at the edge of the ionized region. As the radiation field softens, there is a reduction in the UV photons that heat the gas, resulting in a decrease in the [CII] flux. Since both UV and visible photons can heat the grains, the [CII] to FIR ratio decreases with age.

We do not favor a soft UV field arising from an aging starburst as the explanation for the [CII] deficit. Since starburst luminosities decrease as the stellar population ages, older starbursts in ULIRGs would imply that these same ULIRGs were much more luminous in the past, inconsistent with the observed luminosity distribution of IR galaxies in the local universe. This explanation is therefore not very compelling since it would imply that the less luminous objects are characterized by younger starbursts. S. Satyapal et al. (in preparation) model the [CII] to FIR ratio as a function of starburst age. They show that, if an aging starburst produces the low [CII] to FIR ratio in Arp 220, the starburst taking place in this galaxy would need to be at least $3 \times 10^6$ years older than the one taking place in the less luminous galaxy Arp 299. If the starburst in Arp 220 were 3 $\times 10^6$ years older than the one taking place in Arp 299, the luminosity of Arp 220 would have been roughly three times greater than its current value several million years ago and almost an order of magnitude greater a couple of million years earlier. If ULIRGs are indeed older starbursts, we would expect to see a population of $>10^{13}$ $L_\odot$ ULIRGs in the local universe, which we do not.

### 3.4. Optical Depth and Extinction Effects

In Paper I, we mentioned the possibility of self-absorbed or optically thick [CII] emission. As discussed, self-absorbed [CII] requires that $C^+$ exist in cooler foreground material. Self-absorbed [CII] emission has been observed towards the HII region/molecular cloud complex Sgr B2 near the Galactic Center (Cox et al. 1999; Vastel et al. 2002). In Sgr B2, FIR fine-structure line emission from ionized and PDR gas is extremely faint (Cox et al.), as is the case in ULIRGs, and the [OI] 63 μm line is seen in absorption, as is the case in some ULIRGs such as Arp 220. However, we do not believe that the [CII] deficit observed in ULIRGs is caused primarily by self-absorption. Rather, we suggest and discuss further in the next section, that the cause of the [CII] deficit in these sources is also responsible for the faintness of the [OI] 145 μm line, which we note does not appear to show self-absorption in Sgr B2 (Vastel et al.), as well as that of the [OIII] 52 and 88 μm and [NIII] 57 μm lines. As for optically thick [CII], the $C^+$ column densities needed to attain the required optical depth in the [CII] line, i.e., $\tau_{[CII]} \geq 2$, correspond to an unrealistically high $A_V$ (>130 mag) in the PDR gas alone compared to previous measures of the total $A_V$ for the cold molecular and PDR gas (Sturm et al. 1996).

Comparison of mid-IR recombination and fine-structure line fluxes with the free-free millimeter continuum in the prototypical ULIRG Arp 220 suggests that extinction high enough to produce obscuration at FIR wavelengths, or specifically 158 μm, is not evident (Genzel et al. 1998; S. Satyapal et al., in preparation), at least for the case where the gas responsible for the observed mid-IR recombination and fine-structure line emission is also responsible for the millimeter emission. This result would tend to rule out high extinction at 158 μm as the origin of the [CII] 158 μm deficit. Furthermore, significant extinction at 158 μm is not supported by the relatively normal [OI] 63 μm to [CII] 158 μm line ratios in the ULIRGs in this study and the tendency toward higher [OI] to [CII] ratios in the [CII] deficient galaxies in the study of Malhotra et al. (2001), since high differential extinction would act to decrease the [OI] 63 μm emission relative to [CII] and produce less than normal [OI] 63 μm to [CII] line ratios. For $A_\lambda \propto \lambda^{-\alpha}$ and $1 \leq \alpha \leq 2$, we would need to multiply the observed [OI] 63 μm to [CII] line ratio by a factor of at least 16 (for $\alpha = 1$ and at most $2.5 \times 10^5$ for $\alpha = 2$) to correct for extinction if extinction at 158 μm alone accounts for the [CII] deficit, i.e., if $A_{158\mu m} \sim 2$. Such a correction to the observed, relatively normal [OI] to [CII] ratios would generate flux ratios greatly in excess of the predictions of the PDR models for any reasonable n and $G_0$.

Although extinction at FIR wavelengths cannot explain the [CII] deficit, mid-IR spectroscopic observations by Genzel et al. (1998) do appear to indicate higher mid-IR extinction toward the ionized medium in ULIRGs than in normal and starburst galaxies. Because of uncertainties in the nature and exact shape of the extinction curve in the mid-IR, the degree to which extinction may lower the PAH flux relative to FIR in ULIRGs as compared to normal and starburst galaxies is difficult to quantify. For example, the Draine (1989) mid-IR extinction law ($A_\lambda \propto \lambda^{-1.75}$) implies a factor of only 1.7 drop in the 6.2 μm PAH flux going from $A_V = 10$ (M 82; Genzel et al. 1998) to $A_V = 45$ (Arp 220; Genzel et al. 1998). To explain the PAH deficit relative to FIR in ULIRGs would require $A_V > 100$, in disagreement with the extinction estimates of mid-IR spectroscopy. Lutz (1999), however, derives a mid-IR extinction curve based on ISO observations of the highly-obscured Galactic Center region that indicates significantly higher extinction in the 3 to 8 µm region than implied by Draine. According to Lutz, the 6.2 µm flux should drop by nearly a factor of five going from $A_V = 10$ to $A_V = 45$, in which case, mid-IR extinction could account for much of the PAH deficit if the PAH emitting regions suffer the same extinction as that derived from the recombination and fine-structure lines. However, this may not be the case. For example, recent evidence suggests that the mid-IR continuum in AGN may suffer more extinction than the PAHs. Clavel et al. (2000) find that, although the mean equivalent width of the 7.7 µm feature in Seyfert 1 galaxies is a factor of 5.4 lower than that of their Seyfert 2 counterparts, the mean PAH luminosities of the two types are equal. Clavel et al. attribute the high mean equivalent width in Seyfert 2 galaxies to extinction of the mid-IR continuum. This would imply that in Seyfert 1 and 2 galaxies, the PAH extinction is the same even though the continuum emission suffers very different extinction.

Given the uncertainties in the mid-IR extinction, the relative constancy in the PAH to [CII] ratio between ULIRGs and normal/starburst galaxies, as well as in the Helou et al. (2000) sample, is indeed remarkable. It would therefore be somewhat surprising if effects such as lower photoelectric heating (which would lower the [CII] to FIR flux) and higher average differential extinction (which would lower the PAH to FIR flux) in ULIRGs just cancel each other to produce the observed constant ratio, especially considering the large "moment arm" in wavelength space between the 6.2 µm PAH feature and the [CII] 158 µm line.

**3.5. Dust within Photoionization Environments**

Based on the analysis of section 3.2, we see that the application of PDR models under simplifying assumptions does not yield consistent results. The differences in [CII], CO, [CI], and presumably, FIR beam area filling factors imply that a significant fraction of the FIR could come from environments where there is little or no [CII] or PAH emission. Photoionization regions may provide one such environment if the dust in the ionized gas is a primary contributor to the overall FIR luminosity.

In photoionization regions, dust grains compete with the gas in absorbing the direct stellar and nebular continuum radiation. Dust grains reradiate the absorbed photons as thermal continuum radiation in the IR. We would expect ionized gas with dust grains that compete efficiently with the hydrogen for the available UV photons to exhibit enhanced FIR emission relative to the recombination and fine-structure line emission from the gas. Compared to their less luminous counterparts, the mid- to far-IR spectra of ULIRGS do in fact display a lack of strong fine-structure emission lines from ionized gas, as well as an abundance of molecular absorption lines (Lutz et al. 1996; Genzel et al. 1998; Fischer et al. 1999). Fischer et al. argue that the weakness of the FIR fine-structure lines is not due to FIR extinction. Voit (1992) suggests that direct absorption of the ionizing luminosity by dust could reduce fine-structure and recombination line emission in ULIRGs. The warmer FIR colors and dust temperatures of ULIRGs (e.g., Sanders et al. 1988) compared to less luminous galaxies suggest that the role of dust in ionized gas is important.

As discussed by Voit (1992) and Bottorff et al. (1998), dust absorbs most of the ionizing photons when the dimensionless ionization parameter U is high, where U is defined as:

$$U = Q / (4 \pi r^2 n_e c) = \phi(H) / n_e c = N_{HII} \alpha_B / c$$

Q and $\phi(H)$ are the rate and flux of ionizing photons, $n_e$ is the electron density, c is the speed of light, $N_{HII}$ is the ionized hydrogen column density, and $\alpha_B$ is the case B hydrogen recombination coefficient (Storey & Hummer 1995). Extragalactic clouds with larger ionization parameters have larger ionized column densities and larger grain column densities, which results in the absorption by dust of a larger fraction of the incident luminosity. For clouds with standard Galactic dust abundance and composition, dust absorbs most of the incident ionizing radiation above $U = 10^{-2}$ (Voit 1992; Bottorff et al. 1998). In such cases, the ionized hydrogen column density is typically $10^{21}$ cm$^{-2}$ or greater. Bottorff et al. show that the IR to H$\beta$ ratio is a strong function of U, with the IR intensity increasing as grains absorb more radiation and H$\beta$ decreasing as hydrogen absorbs fewer ionizing photons. Such "dust-bounded" cases fall in the IR / H$\beta$ >> 100 limit (Bottorff et al. 1998). Based on models with the photoionization code CLOUDY (Ferland 1993), Fischer et al. (2001) find that high U conditions may explain the lack of detected fine-structure line emission in ULIRGs. For example, these models predict that for central instantaneous starbursts and power law sources with $Q = 4.5 \times 10^{54}$ s$^{-1}$ and $n_e$ = 500 cm$^{-3}$, lowering the distance between the central source and the surrounding gas from 500 pc (U = $10^{-2}$) to 50 pc (U = $10^0$) decreases the ratio of the [OIII] 88 μm line to the total luminosity by more than an order of magnitude.

Enhanced dust absorption in the ionized gas offers a natural explanation for the implied beam area filling factor differences between the [CII] emission and the CO, [CI], and FIR emission. If a significant portion of the FIR emission emanates from the photoionized gas, whereas the [CII] does not, we would expect this extra source of FIR emission to increase the apparent FIR beam area filling factor relative to that of [CII]. Here again, we emphasize that unless the FIR emission from the ionized gas is taken into account, either by applying a beam area filling factor correction or by subtracting the FIR contribution from the ionized gas from the total FIR emission, the PDR model predictions in section 3.1 do not apply to ULIRGs. By inhibiting the penetration of 13.6–6 eV photons (i.e., those photons thought to be responsible for heating the PDR gas) and decreasing the extent of the [CII]-emitting region, dust-bounded ionized gas should also give rise to [CII] beam area filling factors that are smaller than those for CO and [CI]. This is indirectly supported by numerous extragalactic observations of an effect in the opposite sense, whereby a reduced dust abundance allows the UV radiation to penetrate further into the interstellar medium, enhancing [CII] / CO (1-0) and [CII] / [CI] (1-0) and increasing the beam area filling factor of [CII] relative to CO and [CI] (e.g., Maloney & Black 1988; Israel et al. 1996; Madden et al. 1997; Madden 2000).

The peak wavelength of the dust emission from high U regions will depend on the magnitude of U. Such regions can help explain both the relatively warm 60 / 100 μm colors in ULIRGs (e.g., Sanders et al. 1988) and the weakness of the PAH emission relative to the total FIR emission as traced by the 6.2 μm PAH feature and the IRAS 12 μm flux. Boulanger et al. (1988) show that the 60 μm / 100 μm IRAS band ratio decreases and the 12 μm / 25 μm band ratio increases sharply with distance from the central star within the California nebula HII region. This difference in behavior between the colors within HII regions and in the surrounding PDR is virtually identical to the difference in both the 60 μm / 100 μm and 12 μm / 25 μm colors of ULIRGs compared with less luminous normal and starburst galaxies, highly suggestive of an increase in the contribution from dust in ionized gas to the FIR emission in ULIRGs. In the California nebula, the 60 µm flux is relatively constant throughout the HII region, while the 12 µm emission has a strong peak at the edge of the HII region. These effects are attributed to the destruction of the 12 µm emitters and the presence of large grains throughout, because the emission from large grains gradually shifts toward shorter wavelengths as the equilibrium temperature increases. Mouri, Kawara, & Taniguchi (1997) argue that the mid-IR emission in starburst galaxies does in fact come from large grains within HII regions. Boulanger et al. attribute the destruction of the 12 µm emitters to the increased strength of the radiation field closer to the star. Desert, Boulanger, & Puget (1990) model this behavior in more detail with a three component model but state that the total intensity of the radiation field is less critical a factor than the presence of hard UV photons. Hard UV photons are present within HII regions and, of course, in ionized regions around AGN. On the other hand, in PDR environments, hard UV photons are largely absent, and for the $G_0$ values derived here for ULIRGs, destruction of PAHs is not likely to occur.

Is it reasonable that high U conditions contribute in an energetically important way in ULIRGs, as we suggest? Using mid-IR spectroscopy to estimate the ratio of the ionizing luminosity $L_{Lyc}$ to the bolometric luminosity $L_{bol}$, Genzel et al. (1998) find $L_{Lyc}$ / $L_{bol}$ values in ULIRGs to be nearly half that in their starburst templates. Systematic beam size effects in their starburst templates could possibly make the discrepancy larger by a significant factor. In our scenario, measured Q / L values *would* be significantly lower in ULIRGs if their intrinsic values are similar. Given the uncertainties in the extinction estimates and possible beam size effects in the nearby starburst nuclei, further measurements are warranted.

Why might high U (= $\phi(H) / n_e c$) conditions exist in ULIRGs and what possible configurations might we envisage? Given the evidence for relatively higher densities in ULIRGs, the high U conditions likely arise from high $\phi(H)$ due to high Lyman continuum luminosities impinging on material at small radii, rather than from low electron densities. These conditions could arise in highly compact HII regions, or they could arise from differences in the global conditions in ULIRGs compared with luminous IR galaxies, such as might be present in very compact luminous starburst regions where the ionized regions of large numbers of individual HII regions are merged and possibly surrounded by a global PDR. Here, we might expect high U to be accompanied by high $G_0$.

For the most extreme high U cases, such as might be present in buried or partially obscured AGN, the dust emission at the dust sublimation radius must peak at near-IR wavelengths (Dudley & Wynn-Williams 1997). In this case, radiative transfer effects could produce excess FIR emission arising from a central source that is not associated with the normal PDR tracers in these galaxies and may suffer higher extinction than the PDR components. This explanation may be responsible for the observed FIR molecular absorption seen towards ULIRGs (Fischer et al. 1999). Such entities could be present together with a component of more normal and possibly less obscured PDR and starburst regions. These PDR and starburst regions could dominate the line emission and thus mask the presence of an obscured central source. In this paper, we do not try to ascertain quantitatively which of these scenarios is dominant, although pioneering work along these lines by Laurent et al. (2000) has attempted to quantify separate PDR, HII regions, and AGN contributions to the mid-IR spectral signatures of galaxies. Rather, one of the aims of this paper is to quantify the observational evidence that high U effects are important and to emphasize the need for further observational and modeling efforts.

### 3.6. Implications for More Distant Sources

As discussed in Paper I, an important implication of the [CII] deficit in ULIRGs is that the usefulness of the [CII] line in cosmological work could be diminished if cosmological sources resemble this local sample. As a result of the present work, we can begin to provide an indication of what this means quantitatively. In the present sample, only three of 15 ULIRGs resemble normal and starburst galaxies in the ratio of [CII]-to-dust continuum so that an experiment that could just detect a cosmological ultraluminous IR source, such as those detected in continuum using SCUBA with normal [CII] properties, would need to observe five sources for one [CII] detection. Specifically, for the high-redshift sources considered by Stark (1997), the per season detection rate estimate would be reduced from 200 to 40. Such a detection rate should still provide useful information for understanding star formation as a function of redshift (e.g., Madau,

Pozzetti, & Dickinson 1998), since lower limits to the massive star formation rate can be calculated directly from the measured [CII] intensity. Given the presence of upper limits to the [CII] flux in our sample, it is not possible to say what would be required to detect the entire sample at cosmological distances, but it is clear that should the [CII] deficit hold, it may become a driver for more ambitious future surveys.

In Figure 9, the ratio of [CII] to FIR flux is plotted against 8−1000 µm IR luminosity for both the sample considered here and the Malhotra et al. (2001) sample. For galaxies not listed in Table 2, we take the IR luminosities from the IRAS Bright Galaxy Sample (Soifer et al. 1989) or its extension (Sanders et al. 1995). As a precaution, we have plotted as open symbols sources which are listed as resolved at 25 or 60 µm or both, though we have made use of the LWS data to estimate the FIR flux where appropriate in the present sample. It is of interest that for the strongest upper limit in this plot (NGC 4418), the power source of the FIR is considered to be a buried AGN based on its mid-IR spectral properties (Roche et al. 1986; Dudley & Wynn-Williams 1997; Spoon et al. 2001). The data in Figure 9 suggest a break occurring near $10^{12}$ $L_\odot$, though additional data, say with the Herschel and ASTRO-F space telescopes, are required to rule out a continuing decline at higher luminosities and to better define the sharpness of the break.

Since the [CII] 158 µm line is one of the brightest lines in normal and starburst galaxies (and ULIRGs) in the wavelength range where dust extinction plays a lesser role, its usefulness as an astrophysical diagnostic remains undiminished notwithstanding the cautionary implications of the present work. In the context of the high U hypothesis discussed above, sources with low dust abundance relative to hydrogen would have to achieve an even higher U to produce a similar diminution of the [CII] line intensity relative to the continuum. For such scenarios, we might expect the break suggested in Figure 9 to migrate to higher luminosities if U and L are generically correlated and enrichment is a gradual function of cosmic time in individual sources. If such is the case, then mid- and far-infrared fine-structure lines, which are preferred ISM abundance diagnostics (e.g., Simpson et al. 1998, 1995), as well as the [CII] 158 µm line itself, will be easily observed in moderate-luminosity cosmological ultraluminous IR sources and can provide a check on this scenario. These could be the majority of such sources if the luminosity function slope near $10^{12}$ $L_\odot$ at earlier times is similar to its present value (see, for example, Soifer & Neugebauer 1991; Kim & Sanders 1998).

Alternatively, since ULIRGs are mainly associated with major mergers, it may be that currently known cosmological ultraluminous IR sources will also be associated with (pairs of) well-formed and thus almost certainly pre-enriched systems. If this is the case, the break near $10^{12}$ $L_\odot$ may be intrinsic to the properties involved in the formation processes associated with ULIRGs. In this scenario, migration of the break in Figure 9 may not be observable at sub-millimeter wavelengths. A more detailed discussion of the cosmological implications of the break in the [CII] emission around $10^{12}$ $L_\odot$ in light of the above scenarios can be found in Dudley et al. (2003).

## 4. SUMMARY

In Paper I, we reported the detection of a deficiency in the [CII] 158 μm emission relative to the FIR continuum in a small sample of ULIRGs. In this study, we combine all the available [CII] spectra obtained with ISO to enlarge the sample from seven to 15 ULIRGs, 12 of which show detectable [CII] emission. We find that the enlarged sample of ULIRGs confirms our earlier observation that the [CII] flux in ULIRGs is about an order of magnitude lower than that seen from nearby normal and starburst galaxies. As discussed in Paper I, this result could have important implications for the use of the [CII] line as an eventual tracer of high-redshift protogalaxies, which may resemble the disturbed ULIRGs at low redshift observed in this study.

In an attempt to address the origin of the [CII] deficit, we combine the [CII] data with ISO LWS spectra of the [OI] 63 μm line fluxes and [OI] 145 μm upper limits in the brightest galaxies in our sample; like [CII], the [OI] lines are thought to arise primarily in PDR gas. We have combined the [OI] measurements with other recent observations of the atomic fine-structure and molecular rotation line radiation and PAH emission in ULIRGs to explore the origin of the [CII] deficit. In particular, we have for the first time examined in detail the self-consistency and applicability of the PDR assumptions in the context of a filling factor analysis for many of the most often observed diagnostics.

Based on the [CII] and [OI] observations alone, we find that PDR models suggest that ULIRGs are characterized by high $G_0$ / n PDRs, but only *if the PDR model* *assumption that the [CII], [OI], and FIR emission components are coextensive holds*, i.e., both the lines *and* the FIR trace PDR gas and have identical beam area filling factors. Within the context of these models the observed decrease in [CII] / FIR and 6.2 μm PAH / FIR with 60 μm / 100 μm (which scales with $G_0$ in the PDR) provides apparent support for the presence of high $G_0$ / n PDRs in ULIRGs. We do find, however, that any PDR model that can account for the observed [CII], [OI] 63 and 145 μm, and FIR emission does not compare well with previous observations of the neutral/molecular ISM in ULIRGs.

We use PDR models to derive beam area filling factors for [CII], CO (1-0), and [CI] (1-0). Inconsistencies in the derived beam area filling factors suggest that the assumption that the [CII] and FIR fluxes arise from the same region does not apply for ULIRGs. If true, we argue that non-PDR components that dominate the FIR emission but do not contribute to the [CII] flux can account for the [CII] deficit. (Beam filling factor effects do not rule out an added contribution from high $G_0$ / n PDRs to the [CII] deficit; they only preclude the use of one-dimensional, plane-parallel models as a means to justify a high $G_0$ /n scenario.) We point out that, for previous PDR analyses that assume that the FIR flux arises solely from PDRs, disparate PDR physical conditions are derived depending on the lines used: [CII] and [OI] alone yield high $G_0$ / n solutions, whereas when CO (1-0) and [CI] (1-0) are compared with [CII], high n and high $G_0$ are derived. This discrepancy further suggests that non-PDR contributors to the FIR emission are at work – a finding supported by the observed similarity in the [CII] to 6.2 μm PAH ratio in ULIRGs and less luminous galaxies. Additionally, we conclude that soft UV radiation fields arising from aging starbursts, optical depth and FIR extinction effects, and high n and high $G_0$ ($n > 10^5$ cm$^{-3}$, $G_0 > 10^4$) PDRs are not compatible with the full set of FIR line and continuum observations (although mid-IR extinction may play a role in the PAH deficit relative to FIR).

As an example of a non-PDR FIR component, enhanced dust absorption in the ionized gas may be able to account for the implied beam area filling factor differences between [CII] and CO, [CI], and FIR. Enhanced dust absorption of ionizing photons in ionized regions could give rise to significant FIR emission from the photoionized gas. This same gas would not produce strong [CII] emission. Such an effect could give rise to smaller beam area filling factors for [CII] than for FIR and produce a [CII] deficit relative to FIR as observed. Photoionized gas characterized by a high U can produce such enhanced dust absorption and excess FIR emission. High U ionized regions also provide an explanation that simultaneously accounts for the suppressed FIR fine-structure line and PAH emission and the warmer FIR colors and dust temperatures of ULIRGs compared to less luminous galaxies.

Should the [CII] deficit hold in high-redshift protogalaxies, it could have an impact on the use of the [CII] line for future surveys. An estimate of the detection rate of such objects based on the [CII] observations presented here would imply, however, that future [CII] observations may still yield useful information for understanding star formation as a function of redshift. Here, the dust abundance of these high-redshift objects will play an important role, since sources with low dust abundance relative to hydrogen would have to achieve an even higher U to produce a [CII] deficit similar to that seen for the ULIRGs. Future observations of the [CII] line in high-redshift sources may therefore help to provide a check on the proposed high U scenario in ULIRGs.

We are grateful to our collaborators on the LWS IR-bright galaxies science team P. Clegg, J. Colbert, P. Cox, M. A. Greenhouse, V. Harvey, C. M. Bradford, S. Lord, M. Malkan, G. Melnick, H. Smith, L. Spinoglio, G. Stacey, and S. Unger, for their contributions to the IRBGALS program. We would also like to thank the LWS instrument and data analysis teams at Vilspa, RAL, and IPAC. This work was supported by the Office of Naval Research, the NASA ISO grant program, the Presidential Early Career Award for Scientists and Engineers program (SS), and the NASA Long Term Space Astrophysics grants NAG5-9271 (MWG) and S-92521-F (JF & CD). The ISO Spectrometer Data Center at MPE is supported by DLR under grants 50 QI 8610 8 and 50 QI 9402 3.

**APPENDIX**

In this appendix, we discuss the basis for the assumption in this work that the [OI] and [CII] lines in extragalactic sources are optically thick and thus arise from the near side of the clouds, as adopted in the Kaufman et al. (1999) PDR models. For plane parallel models, this produces a factor of 0.5 in the ratios of these line fluxes to the total FIR flux as discussed in section 3.1.

Kaufman et al. model a typical galactic PDR associated with a star forming region in which the line width is 1.5 km s$^{-1}$ and $A_v \geq 10$, corresponding to a total column $\geq 2 \times 10^{22}$ cm$^{-2}$. This analysis can be applied to galaxies as long as most PDRs along the same line of sight have non-overlapping projected velocities. Of the total column, the [OI] emitting region on the far side of a cloud corresponds to approximately $A_v = 3$, or a column of N = N(HI) + 2N(H$_2$) = $6 \times 10^{21}$ cm$^{-2}$. Thus, the [OI] line emission passes through a column of N = $1.4 \times 10^{22}$ (from the far side of the cloud to the observer). In the cloud center, of extent approximately $A_v = 4$, the O is partly tied up in CO, with the remainder mainly in the form of atomic O. From Table 4 of Tielens & Hollenbach (1985), an optical depth of 1 occurs in the [OI] line for a column of only N = $7.8 \times 10^{20}$ cm$^{-2}$. This column should be revised for the differences in gas phase abundance ($5 \times 10^{-4}$ and $3 \times 10^{-4}$) and line width (2.7 and 1.5 km s$^{-1}$) between that of Tielens & Hollenbach and Kaufman et al. (1999) respectively, using only the atomic O which is not tied up in CO ($1.6 \times 10^{-4}$) in the central region. Taking these differences into account, estimates of the optical depth in the [OI] line corresponding to these near side and central parts of the column are:

Near side ($A_v = 3$):
$$\tau[OI]_{63\mu m} = (6 \times 10^{21} \text{ cm}^{-2} / 7.8 \times 10^{20} \text{ cm}^{-2})(3 \times 10^{-4} / 5 \times 10^{-4})(2.7 \text{ km s}^{-1} / 1.5 \text{ km s}^{-1}) =$$
$$8.3$$

Central region ($A_v = 4$):
$$\tau[OI]_{63\mu m} = (8 \times 10^{21} \text{ cm}^{-2} / 7.8 \times 10^{20} \text{ cm}^{-2})(1.6 \times 10^{-4} / 5 \times 10^{-4})(2.7 \text{ km s}^{-1} / 1.5 \text{ km s}^{-1}) =$$
$$5.9$$

This yields a total optical depth of $\tau[OI]_{63\mu m} = 14.2$. With an optical depth for [OI] 63 μm on the order of 14, a fraction of only $e^{-14} \approx 10^{-6}$ of the emission penetrates from the far-side through the near-side of the cloud. (We note that the fact that self-absorption is rarely seen means that there is little cold gas between the last emitting region, the near side of the cloud, and us, not that the line is optically thin. In addition, the temperature structure of the PDR, with hotter regions near the surface, and cooler regions in the interior, is not expected to produce self-absorbed profiles even for optically thick lines emitted from the near side of the cloud.)

For $C^+$, Tielens & Hollenbach (1985) calculate an optical depth of 1 for $C^+$ at a column density of $1.2 \times 10^{21}$ cm$^{-2}$. This column density should be revised by a factor of $(3.0 \times 10^{-4} / 1.4 \times 10^{-4})(1.5$ km s$^{-1}$ / 2.7 km s$^{-1}) = 1.2$ to yield a column of $1.4 \times 10^{21}$ cm$^{-2}$ required to produce an optical depth of 1. The $C^+$ column extends for an $A_v$ of about 2 into the PDR, but the optical depth is dependent upon $G_0$ and n (and T). An $A_v$ of 2 would give an optical depth in the line of about 2.8. From our detailed PDR models, at the typical values $G_0 = 10^3$ and n = $10^3$, we find an optical depth of about 1 in [CII] 158 μm. Thus, we expect that less than 40% of the [CII] emission can penetrate from the far side of the cloud through the near side to us, i.e., the [CII] emission is marginally thick. We note that the fact that [CII] absorption is seen towards the Galactic Center at some velocities (Vastel et al. 2002) arises from a unique perspective in which we view many clouds that are predominantly moving tangentially to our line of sight. This produces a coherent path length of [CII] emission, which leads to self absorption. The same perspective does not generally apply when we view an external galaxy.

TABLE 1

Observing Log

| Source (1) | $\alpha_{2000}$ (2) | $\delta_{2000}$ (3) | ISO Program (4) | AOT (5) | Observation Date (6) |
|---|---|---|---|---|---|
| IRAS 05189-2524 | 05 21 01.4 | -25 21 44.9 | CII_ULIRG | LWS02 | 3 Apr 1998 |
| IRAS 12071-0444 | 12 09 45.3 | -05 01 13.3 | IRBGALS | LWS02 | 19 Jul 1996 |
| Mkn 231 | 12 56 14.2 | +56 52 24.9 | LWS_PV1 | LWS01 | 07 Jan 1996 |
|  | 12 56 14.2 | +56 52 24.9 | IRBGALS | LWS01 | 15 May 1996 |
|  | 12 56 14.2 | +56 52 24.9 | IRBGALS | LWS01 | 11 Jul 1997 |
| Mkn 273 | 13 44 42.1 | +55 53 13.2 | MPEXGAL2 | LWS02 | 25 Apr 1996 |
|  | 13 44 42.1 | +55 53 13.2 | IRBGALS | LWS02 | 19 May 1996 |
|  | 13 44 41.6 | +55 53 18.7 | OHMEGA | LWS01 | 22 Jul 1997 |
| IRAS 15206+3342 | 15 22 38.1 | +33 31 33.3 | IRBGALS | LWS02 | 5 Sep 1997 |
| IRAS 15250+3609 | 15 26 59.3 | +35 58 37.2 | IRBGALS | LWS02 | 20 Sep 1996 |
| Arp 220 | 15 34 57.2 | +23 30 11.2 | IRBGALS | LWS01 | 20 Aug 1996 |
|  | 15 34 57.2 | +23 30 11.2 | IRBGALS | LWS02 | 18 Jul 1997 |
|  | 15 34 57.2 | +23 30 11.2 | IRB_HD | LWS01 | 17 Aug 1997 |
| IRAS 17208-0014 | 17 23 21.9 | -00 17 01.0 | MPEXGAL2 | LWS02 | 5 Mar 1996 |
|  | 17 23 21.9 | -00 17 01.0 | OHMEGA | LWS01 | 26 Aug 1997 |
| IRAS 19254-7245 | 19 31 21.6 | -72 39 24.9 | MPEXGAL2 | LWS02 | 11 Sep 1996 |
| IRAS 19297-0406 | 19 32 22.1 | -04 00 02.2 | CII_ULIRG | LWS02 | 21 Mar 1998 |
| IRAS 20100-4156 | 20 13 29.8 | -41 47 34.7 | OHMEGA | LWS02 | 9 Nov 1996 |
|  | 20 13 29.8 | -41 47 34.7 | CII_ULIRG | LWS02 | 21 Mar 1998 |
|  | 20 13 29.8 | -41 47 34.7 | IRBGALS2 | LWS02 | 23 Mar 1998 |
| IRAS 20551-4250 | 20 58 26.9 | -42 39 06.2 | MPEXGAL2 | LWS02 | 20 Apr 1996 |
|  | 20 58 26.9 | -42 39 06.2 | CII_ULIRG | LWS02 | 9 Nov 1997 |
| IRAS 22491-1808 | 22 51 49.3 | -17 52 24.1 | IRBGALS | LWS02 | 21 May 1996 |
| IRAS 23128-5919 | 23 15 47.0 | -59 03 16.9 | MPEXGAL2 | LWS02 | 20 Apr 1996 |
|  | 23 15 47.0 | -59 03 16.9 | CII_ULIRG | LWS02 | 27 Nov 1997 |
| IRAS 23365+3604 | 23 39 01.2 | +36 21 10.0 | SF_GLX_A | LWS02 | 6 Dec 1997 |
| | 23 39 01.2 | +36 21 10.0 | CII_ULIRG | LWS02 | 7 Dec 1997 |

Note – Col. (1): source name; col. (2): right ascension, epoch 2000, units in hours, minutes, and seconds; col. (3): declination, epoch 2000, units in degrees, arcminutes, and arcseconds; col. (4): ISO program name; col. (5): AOT = Astronomical Observation Template, LWS01 = medium-resolution ($R \sim 200$) spectrum covering the full 43-197 μm range of the LWS, LWS02 = medium-resolution line spectrum (3.4 μm width in the case of the [CII] line); col. (6): ISO observation date.

TABLE 2

Sample Properties and Measured Line Fluxes

| Source (1) | D (Mpc) (2) | $\log(L_{IR}/L_\odot)$ (3) | $F_{FIR}$ (4) | $F_{[CII]\,158}$ (5) | $F_{[OI]\,63}$ (6) | $F_{[OI]\,145}$ (7) | $F_{[CII]\,158}/F_{FIR}$ (8) |
|---|---|---|---|---|---|---|---|
| IRAS 05189-2524 | 170 | 12.10 | 5870 | $1.33 \pm 0.16$ | | | $2.27 \times 10^{-4}$ |
| IRAS 12071-0444 | 514 | 12.29 | 1110 | <0.6 | | | $<5.41 \times 10^{-4}$ |
| Mkn 231 | 169 | 12.54 | 14200 | $3.51 \pm 0.14$ | $2.89 \pm 0.53$ | <0.42 | $2.47 \times 10^{-4}$ |
| Mkn 273 | 151 | 12.11 | 9760 | $5.24 \pm 0.16$ | $4.70 \pm 0.67$ | <0.30 | $5.37 \times 10^{-4}$ |
| IRAS 15206+3342 | 498 | 12.18 | 812 | $1.87 \pm 0.17$ | | | $2.30 \times 10^{-3}$ |
| IRAS 15250+3609 | 222 | 12.00 | 3110 | <1.6 | | | $<5.14 \times 10^{-4}$ |
| Arp 220 | 73 | 12.19 | 47900 | $9.99 \pm 0.24$ | $-6.35 \pm 0.73^{*}$ | <0.94 | $2.09 \times 10^{-4}$ |
| IRAS 17208-0014 | 171 | 12.32 | 14500 | $6.86 \pm 0.18$ | $6.53 \pm 0.96$ | <0.60 | $4.73 \times 10^{-4}$ |
| IRAS 19254-7245 | 247 | 11.99 | 2550 | $2.71 \pm 0.33$ | <2.5 | | $1.06 \times 10^{-3}$ |
| IRAS 19297-0406 | 343 | 12.36 | 3270 | $2.23 \pm 0.11$ | | | $6.82 \times 10^{-4}$ |
| IRAS 20100-4156 | 518 | 12.56 | 2350 | $0.65 \pm 0.09$ | <4.0 | | $2.77 \times 10^{-4}$ |
| IRAS 20551-4250 | 171 | 12.00 | 5410 | $3.51 \pm 0.13$ | $5 \pm 1$ | | $6.54 \times 10^{-4}$ |
| IRAS 22491-1808 | 311 | 12.10 | 2320 | <1.1 | | | $<4.74 \times 10^{-4}$ |
| IRAS 23128-5919 | 178 | 11.95 | 4870 | $5.58 \pm 0.26$ | <5.0 | | $1.15 \times 10^{-3}$ |
| IRAS 23365+3604 | 258 | 12.13 | 3360 | $1.32 \pm 0.10$ | $1.26 \pm 0.39$ | | $3.93 \times 10^{-4}$ |

Note − Col. (1): source name; col. (2): distance for $H_o = 75$ km s$^{-1}$ Mpc$^{-1}$ and $q_o = 0.5$; col. (3): 8−1000 μm luminosity as measured by IRAS using the prescription given in Table 1 of Sanders & Mirabel (1996) [L(8−1000 μm) = $5.6 \times 10^5$ D$_{Mpc}^2$ (13.48 S$_{12}$ + 5.16 S$_{25}$ + 2.58 S$_{60}$ + S$_{100}$) where S is the flux density in Jy]; col. (4): 42.5 to 122.5 μm emission in units of $10^{-13}$ erg cm$^{-2}$ s$^{-1}$ as measured by IRAS Faint Source Catalog using the prescription given in Table 1 of Sanders & Mirabel (1996) [F$_{FIR}$ = $1.26 \times 10^{-11}$ (2.58 S$_{60}$ + S$_{100}$) erg cm$^{-2}$ s$^{-1}$] or as measured by ISO LWS using ISAP (see text for details); cols. (5)-(7): [CII] 158 μm, [OI] 63 μm, and [OI] 145 μm line fluxes in units of $10^{-13}$ erg cm$^{-2}$ s$^{-1}$. Flux upper limits were calculated assuming a Gaussian line with the effective instrumental profile and a 3σ amplitude; col. (8): flux ratio of the [CII] 158 μm line to FIR continuum emission.

$^{*}$The line was seen in absorption, as denoted by the minus sign.

TABLE 3

[CII] Beam Area Filling Factors $\Phi_{[CII]}$ for Four ULIRGs and Comparison Normal/Starburst Galaxies

| Source | Observed $I_{[CII]\,158}$ (erg cm$^{-2}$ s$^{-1}$ sr$^{-1}$) | Predicted $I_{[CII]\,158}$ for $\Phi_{[CII]} = 1$ (erg cm$^{-2}$ s$^{-1}$ sr$^{-1}$) | $\Phi_{[CII]}$ |
|---|---|---|---|
| *ULIRGs* | | | |
| Mkn 273 | $6.0 \times 10^{-6}$ | $2 \times 10^{-4}$ | 0.030 |
| IRAS 17208-0014 | $7.8 \times 10^{-6}$ | $2 \times 10^{-4}$ | 0.039 |
| IRAS 20551-4250 | $4.0 \times 10^{-6}$ | $3.5 \times 10^{-4}$ | 0.011 |
| IRAS 23365+3604 | $1.5 \times 10^{-6}$ | $2 \times 10^{-4}$ | 0.0075 |
| *Normals/Starbursts* | | | |
| M82 | $1.1 \times 10^{-3}$[a] | $4 \times 10^{-4}$ | 2.8 |
| Cen A | $3.3 \times 10^{-4}$[b] | $2.5 \times 10^{-4}$ | 1.3 |
| NGC 1068 | $2.5 \times 10^{-4}$[c] | $2.5 \times 10^{-4}$ | 1.0 |
| NGC 1313 | $1.4 \times 10^{-5}$[d] | $3.5 \times 10^{-4}$ | 0.04 |
| NGC 3256 | $3.8 \times 10^{-4}$[e] | $3.0 \times 10^{-4}$ | 1.3 |
| NGC 3690 (Arp 299) | $1.0 \times 10^{-4}$[f] | $3.0 \times 10^{-4}$ | 0.33 |

[a]Colbert et al. 1999.

[b]Unger et al. 2000.

[c]L. Spinoglio et al. (in preparation).

[d]Contursi et al. 2002 (Region 90).

[e]Carral et al. 1994.

[f]Satyapal et al. 2003.

TABLE 4

CO Beam Area Filling Factors $\Phi_{CO}$ for Four ULIRGs and Comparison Normal/Starburst Galaxies

| Source | Observed $I_{CO\,(1-0)}$ (erg cm$^{-2}$ s$^{-1}$ sr$^{-1}$)[a] | Predicted $I_{CO\,(1-0)}$ for $\Phi_{CO} = 1$ (erg cm$^{-2}$ s$^{-1}$ sr$^{-1}$) | $\Phi_{CO}$ | $\Phi_{CO}/\Phi_{[CII]}$ |
|---|---|---|---|---|
| *ULIRGs* | | | | |
| Mkn 273 | $2.4 \times 10^{-9}$[b] | $2 \times 10^{-8}$ | 0.12 | 4.0 |
| IRAS 17208-0014 | $4.6 \times 10^{-9}$[c] | $2 \times 10^{-8}$ | 0.23 | 5.9 |
| IRAS 20551-4250 | $2.0 \times 10^{-9}$[c] | $4.5 \times 10^{-8}$ | 0.044 | 4.0 |
| IRAS 23365+3604 | $1.7 \times 10^{-9}$[d] | $2 \times 10^{-8}$ | 0.085 | 11 |
| *Normals/Starbursts* | | | | |
| M82 | $1.7 \times 10^{-7}$[e] | $6.5 \times 10^{-8}$ | 2.6 | 0.9 |
| Cen A | $7.6 \times 10^{-8}$[e] | $5.0 \times 10^{-8}$ | 1.5 | 1.2 |
| NGC 1068 | $5.8 \times 10^{-8}$[e] | $5.3 \times 10^{-8}$ | 1.1 | 1.1 |
| NGC 1313 | $4.1 \times 10^{-9}$[f] | $7.5 \times 10^{-8}$ | 0.05 | 1.3 |
| NGC 3256 | $1.0 \times 10^{-7}$[g] | $5.5 \times 10^{-8}$ | 1.8 | 1.4 |
| Arp 299 | $1.5 \times 10^{-8}$[h] | $4.0 \times 10^{-8}$ | 0.37 | 1.1 |

[a]The CO intensities are scaled to the ISO [CII] beam size (69″). For galaxies small compared to the beam size, we divide the observed CO intensity by $8.8 \times 10^{-8}$ sr (69″); for galaxies larger than the beam size, we multiply the observed CO intensity by the ratio of the ISO beam size to the beam size of the CO observations.
[b]Sanders, Scoville, & Soifer 1991 (55″ beam).
[c]Mirabel et al. 1990 (44″ beam).
[d]Solomon et al. 1997 (22″ beam).
[e]Stacey et al. 1991 (55″ beam).
[f]Contorsi et al. 2002 (43″ beam).
[g]Aalto et al. 1995 (43″ beam).
[h]Sanders & Mirabel 1985 (60″ beam).

Figure 1. The [CII] 158 μm line flux observed with the LWS versus the FIR flux for 15 ULIRGs compared with a sample of normal and starburst galaxies (Paper I; Colbert et al. 1999; Stacey et al. 1999; Bradford et al. 1999; J. Fischer et al., in preparation). In the symbol key, the ULIRGs are listed in order of luminosity from top to bottom and are denoted by black filled symbols, while the lower luminosity galaxies are unfilled. The dashed lines mark the regime typical of normal and starburst galaxies.

Figure 2. The 6.2 μm PAH feature versus [CII] 158 μm for ULIRGs (black, filled symbols) and normal/starburst galaxies (open or grey symbols). The galaxy sample is consistent with that in Figure 1 (see text for details). The 6.2 μm PAH emission was measured from ISO CAM CVF spectra (open symbols), integrated over the full LWS beam profile for galaxies with extended emission, and from ISO PHOT-S spectra (filled symbols) for compact galaxies (C. C. Dudley et al., in preparation). Statistical errors bars for the ISO CAM data are shown, although in virtually all cases, the error bars are smaller than the data symbols. Dashed lines show a range of a factor of ten.

Figure 3. The 6.2 μm PAH to [CII] 158 μm line ratio distribution for the galaxies from Figure 2. Upper limits (3σ) on the ratio for two ULIRGs are plotted separately and marked with arrows. The upper limit for IRAS 20100-4156 is 26, well outside the plot boundaries and not a useful contraint to the ratio. Also, we do not plot galaxies from Figure 1 with AGN signatures (NGC 6240, NGC 4945, NGC 1068, and Circinus) in order to compare the ULIRG distribution with that of normal and starburst galaxies only.

Figure 4. Plot of the [C II] 158 μm / [O I] 63 μm ratio versus the ratio of ([C II] 158 μm + [O I] 63 μm) / $I_{FIR}$ emitted from the surface of a PDR. Models of Kaufman et al. (1999) are used with the standard model parameter set and assume that $I_{FIR} = 2 \times [2 \times 1.3 \times 10^{-4}] \, G_0$ erg cm$^{-2}$ s$^{-1}$ sr$^{-1}$. The five ULIRGs − Mkn 231, Mkn 273, IRAS 17208-0014, IRAS 20551-4250, and IRAS 23365+3604 − from this study with both [CII] 158 μm and [OI] 63 μm emission line measurements are shown (filled symbols). Derived

propagation errors resulting from errors in the line and FIR flux measurements imply variations in the depicted values of n and $G_0$ of no more than a factor of a few. For comparison, we plot the objects (open circles) in the Malhotra et al. (2001) sample of 60 normal and starburst galaxies. For one of the points of the Malhotra et al. sample, we depict sample error bars, which are constant for all points. As with the ULIRG sample, we plot only those objects from the Malhotra et al. sample for which both [CII] 158 μm and [OI] 63 μm lines were detected. For Mkn 231, the vector depicts the shift in the location of this galaxy in the figure if we correct for the maximum allowable decrease in the [OI] 63 μm emission owing to self-absorption.

Figure 5. The [CII] 158 μm line to FIR flux ratio versus the IRAS 60 to 100 μm flux ratio for all galaxies plotted in Figure 1. The filled circles denote the ULIRGs, the squares denote the comparison normal and starburst galaxies, and the grey-filled squares denote the galaxies among the comparison sample with AGN-signatures. The error bars for the ULIRGs are statistical; for the comparison galaxies, the statistical errors bars are smaller than the symbols and are therefore not plotted. The arrows represent 3σ upper limits. For extended galaxies, the 60 to 100 μm ratios were derived from the Rice et al. (1988) IRAS catalog rather than the IRAS Faint Source Catalog. For galaxies with no IRAS data (e.g., Circinus), the 60 to 100 μm flux ratio was derived from the LWS full spectrum using the photometry routine in ISAP.

Figure 6. Ratio of the [CII] 158 μm line intensity to the [OI] 145 μm line intensity emitted from the surface of a PDR as a function of the cloud density n and the UV flux incident on the cloud $G_0$. Models of Kaufman et al. (1999) are used with the standard model parameter set.

Figure 7. Ratio of the [CII] 158 μm line intensity to the FIR continuum intensity emitted from the ensemble of extragalactic PDRs as a function of the cloud density n and the UV flux incident on the cloud $G_0$. Models of Kaufman et al. (1999) are used with the standard model parameter set.

Figure 8. The 6.2 µm PAH to [CII] 158 µm line flux ratio versus the IRAS 60 to 100 µm flux ratio for all galaxies plotted in Figure 2. The filled circles denote the ULIRGs, the squares denote the comparison normal and starburst galaxies, and the grey-filled squares denote the galaxies among the comparison sample with AGN-signatures. The error bars and upper limits are 1σ and 3σ values, respectively. The 60 to 100 µm ratios are derived as in Figure 5.

Figure 9. The ratio of [CII] intensity to FIR flux is plotted as a function of IR luminosity for the present sample (circles) as well as for those galaxies in the Malhotra et al. (2001) sample (diamonds) that appear in the IRAS Bright Galaxy Sample. Sources plotted as open symbols may have both continuum and line emission that extend beyond the LWS beam (see text). The sample medians calculated (ignoring upper limits) for the filled diamond symbol galaxies and for the filled circle galaxies with $L_{IR}$ greater than or equal to $10^{12}$ $L_\odot$ are plotted as solid lines.

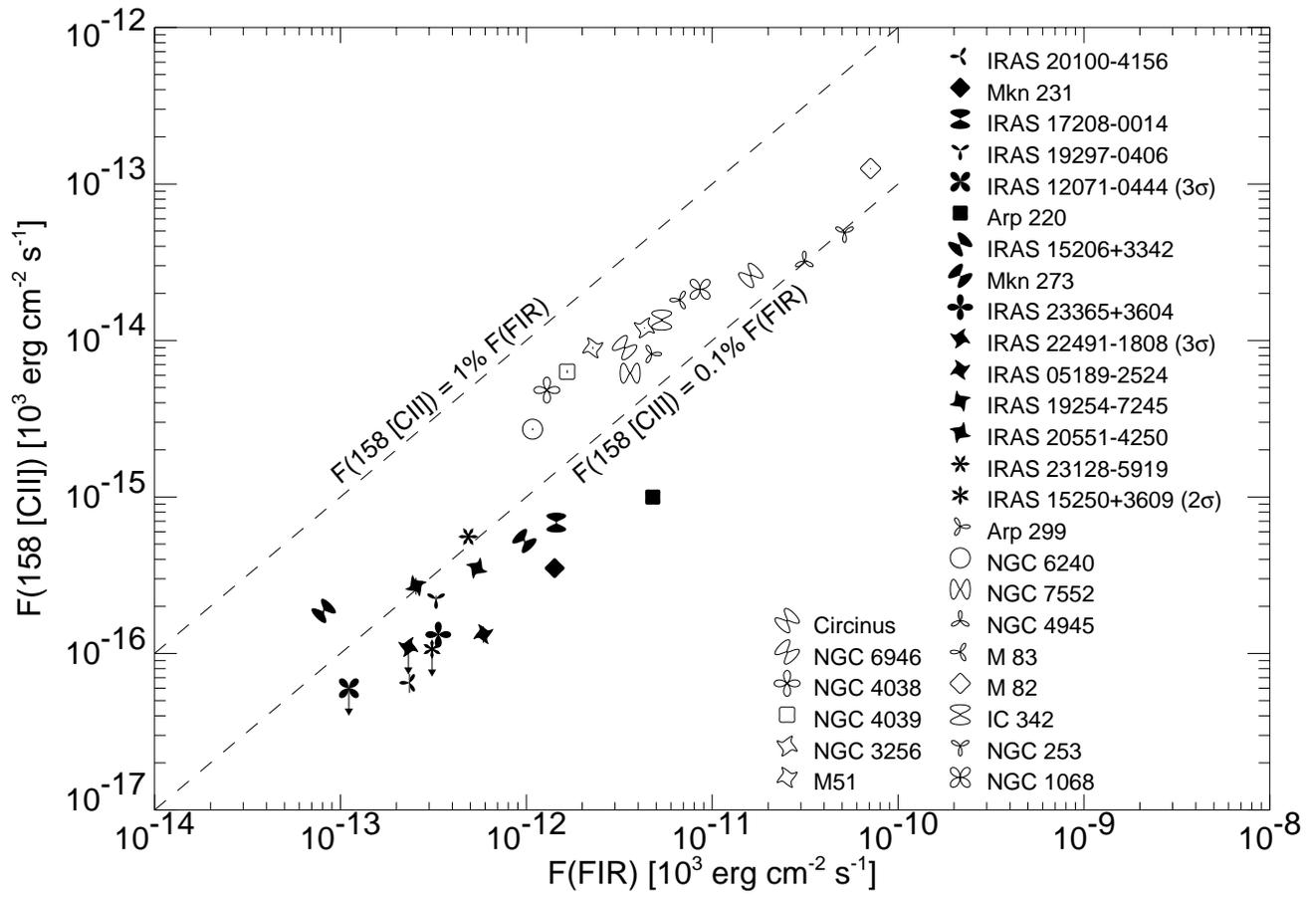

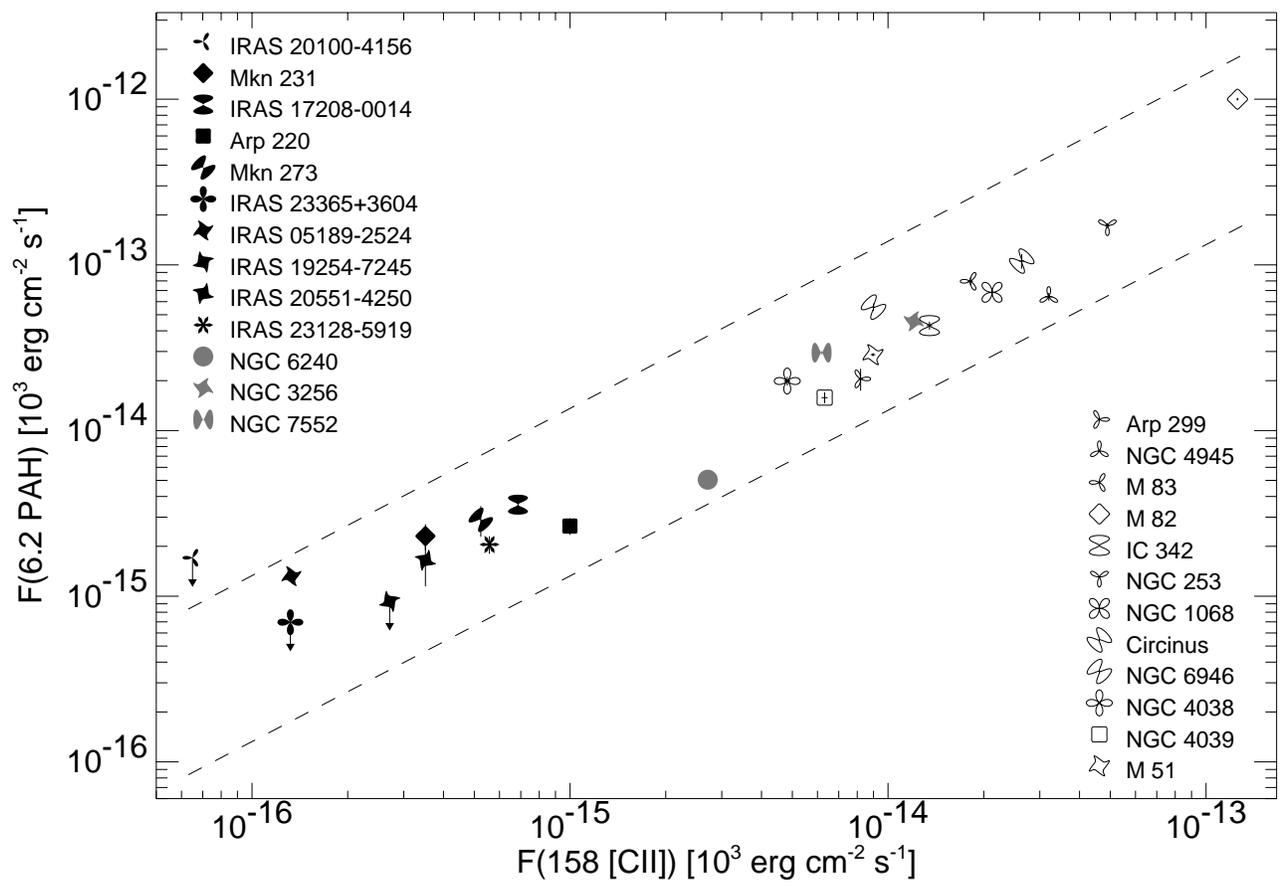

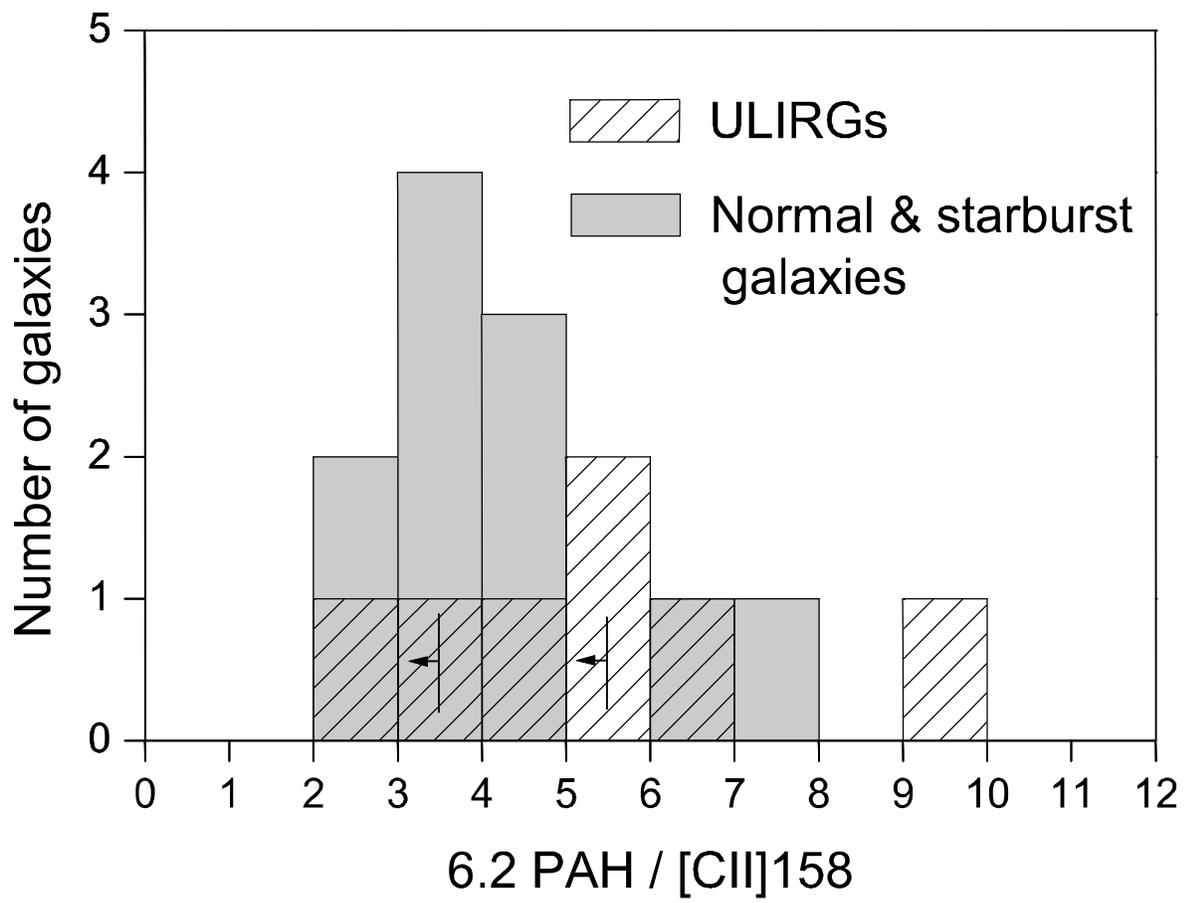

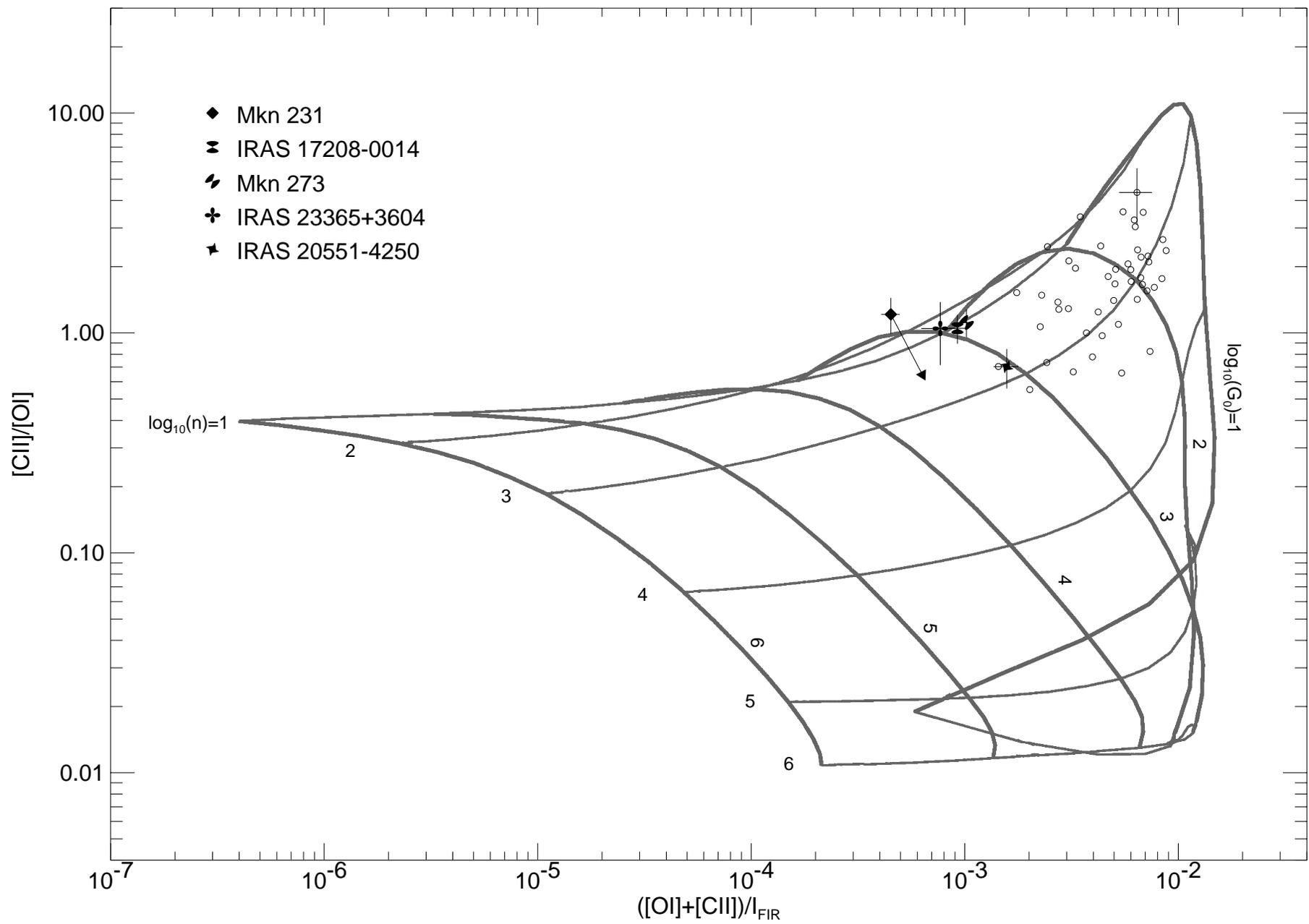

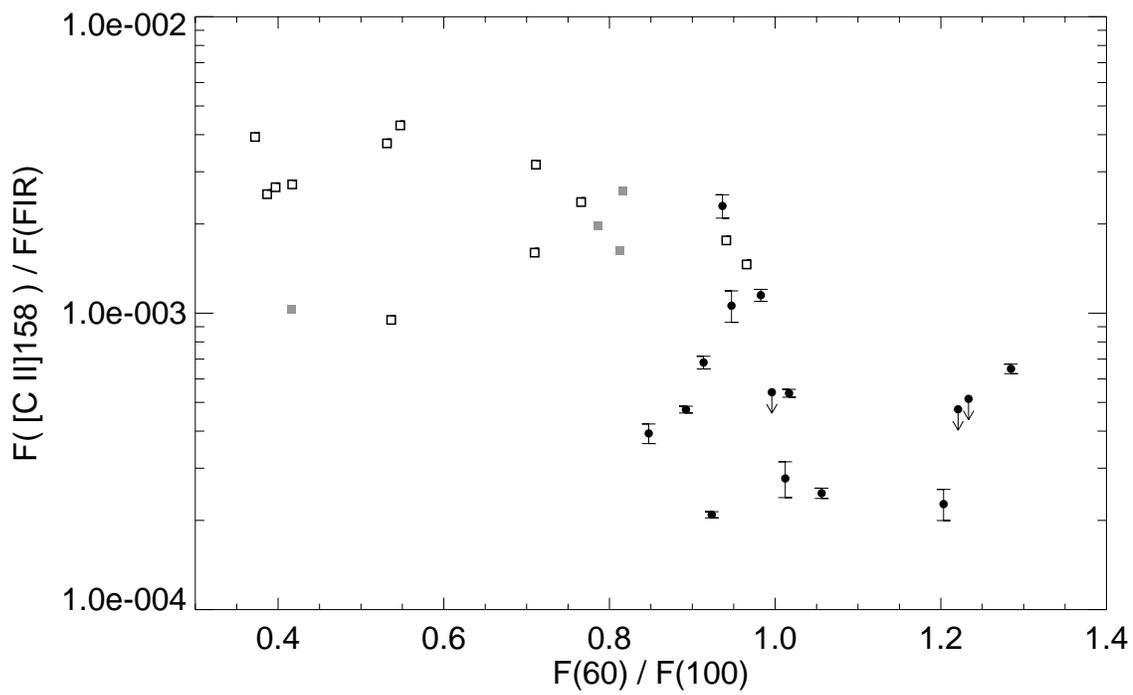

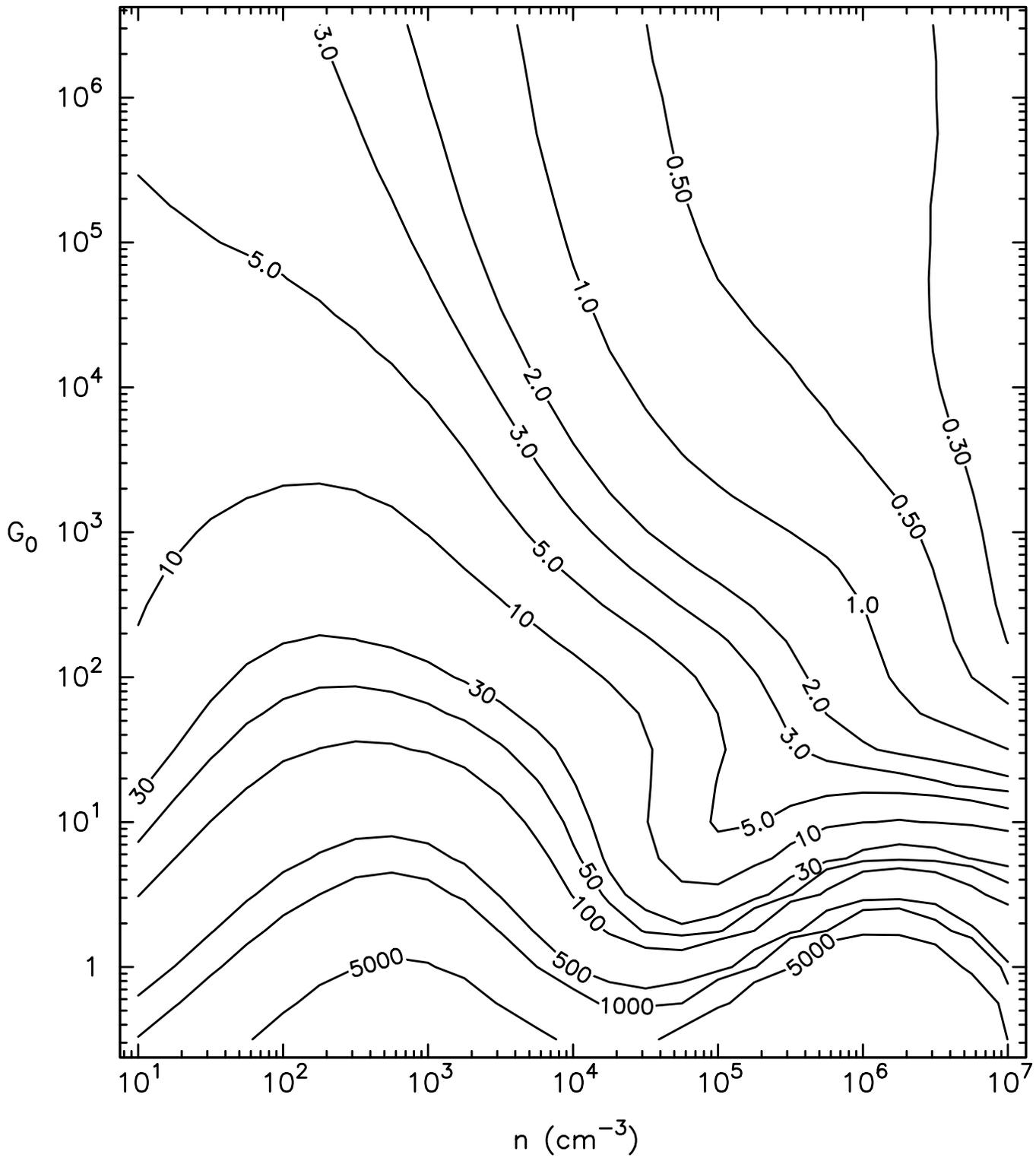

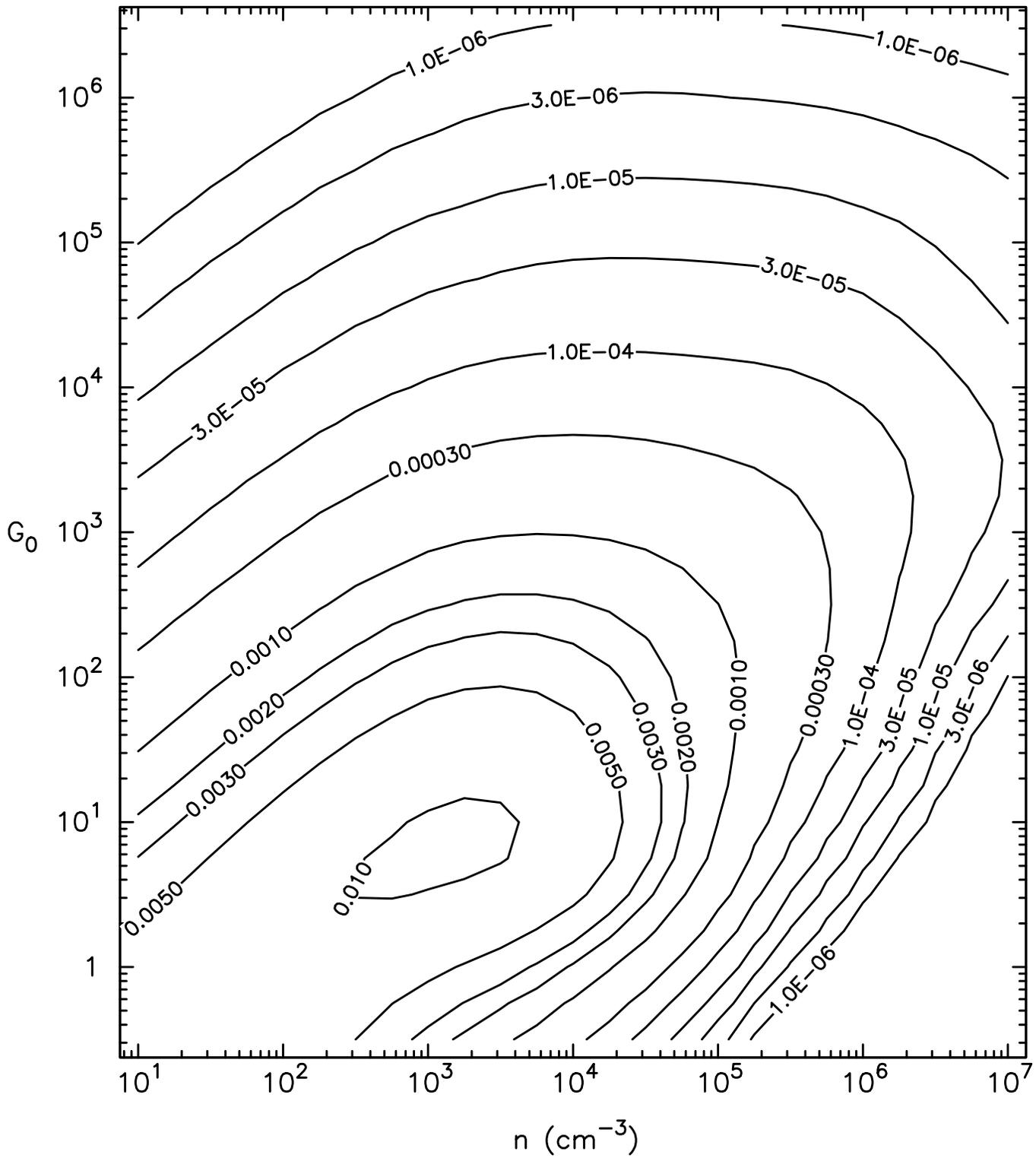

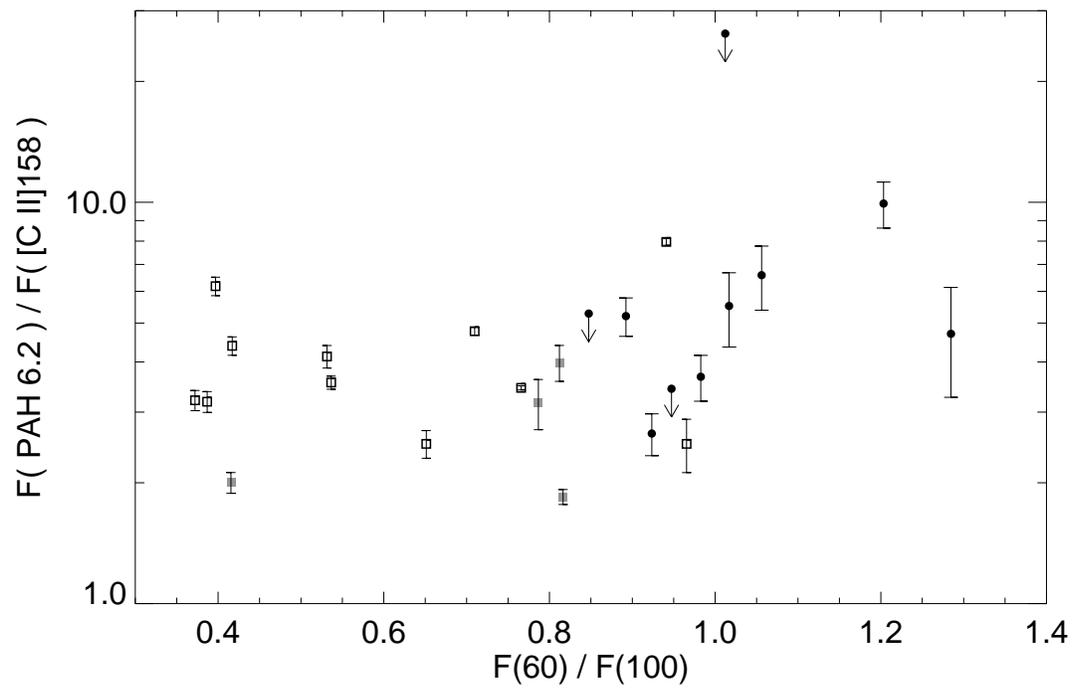

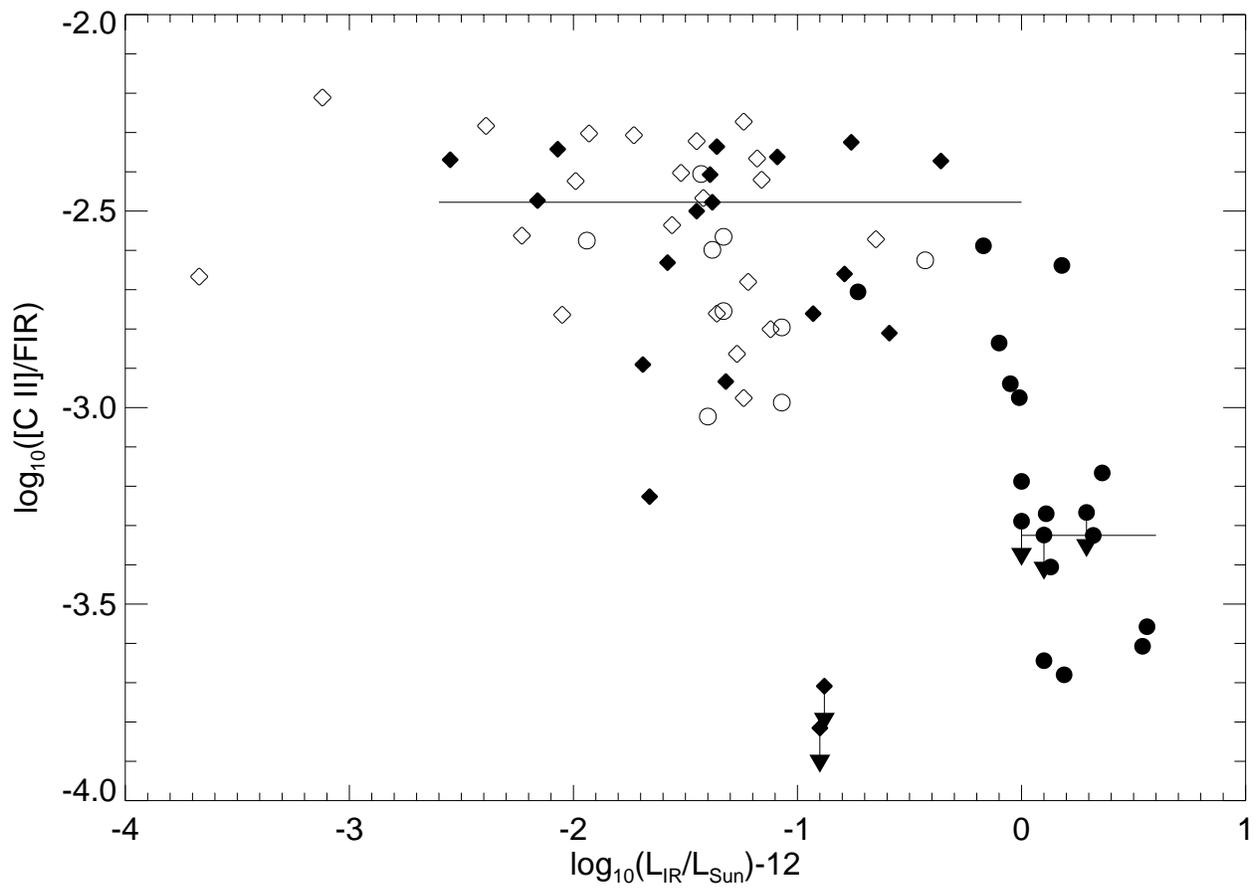